\documentclass[twocolumn,journal]{IEEEtran}
\usepackage{graphicx}
\usepackage{amsmath,amsfonts}
\usepackage{algorithmic}
\usepackage{algorithm}
\usepackage{array}
\usepackage{textcomp}
\usepackage{stfloats}
\usepackage{url}
\usepackage{verbatim}
\usepackage{graphicx}
\usepackage{cite}
\usepackage{subfigure} 
\usepackage{amsmath,amssymb,amsfonts}
\usepackage{algorithmic}
\usepackage{graphicx}
\usepackage{xcolor}
\usepackage{balance}
\usepackage{amsmath}
\usepackage{algorithm}
\usepackage{subcaption}
\usepackage{caption}
\usepackage{amsthm}

\newtheorem{Remark}{Remark}
\newtheorem{proposition}{Proposition}

\usepackage{stfloats}
\newenvironment{proov} {{\it \noindent Proof. }} {\hfill $\blacksquare$\par}

\begin{document}
\title{B-ISAC: Backscatter Integrated Sensing and Communication for IoE Applications}
\author{\vspace{0mm} \large{Zongyao Zhao, {\em Graduate Student Member, IEEE}, Yuhan Dong, {\em Senior Member, IEEE}, Tiankuo Wei, Xinke Tang, Xiao-Ping Zhang, {\em Fellow, IEEE}, and Zhenyu Liu, {\em Member, IEEE}}.

\thanks{Part of this work has been accepted to present at the 2024 IEEE Global Communication Conference (IEEE GLOBECOM 2024), arXiv:2409.0279 \cite{Zhao20241}.}

\thanks{This work was supported in part by the National Natural Science Foundation of China under Grant 62388102, the GuangDong Basic and Applied Basic Research Foundation under Grant 2022A1515010209, and the Shenzhen Natural Science Foundation under Grant JCYJ20200109143016563.\emph{(Corresponding authors: Yuhan Dong, Xinke Tang, and Zhenyu Liu)}}

\thanks{Z. Zhao, Y. Dong and T. Wei are with the Shenzhen International Graduate School, Tsinghua University, Shenzhen 518055, China, and also with the Pengcheng Laboratory, Shenzhen 518000, China (Emails: zhaozong21@mails.tsinghua.edu.cn, dongyuhan@sz.tsinghua.edu.cn, and wtk23@\\mails.tsinghua.edu.cn).}

\thanks{Z. Liu and X.-P. Zhang are with the Shenzhen International Graduate School, Tsinghua University, Shenzhen 518055, China, (Emails: zhenyuliu@sz.tsinghua.edu.cn, xiaoping.zhang@sz.tsinghua.edu.cn).}

\thanks{X. Tang is with the Pengcheng Laboratory, Shenzhen 518000, China (E-mail: tangxk@pcl.ac.cn).}
}

\maketitle
\begin{abstract}
The integration of backscatter communication (BackCom) technology with integrated sensing and communication (ISAC) technology not only enhances the system sensing performance, but also enables low-power information transmission. This is expected to provide a new paradigm for communication and sensing in internet of everything (IoE) applications. In this paper, we propose a novel cognitive wireless system called backscatter-ISAC (B-ISAC) and develop a joint beamforming framework for different stages (task modes). This system can achieve cognitive spectrum sharing between legacy communication, backscatter communication and sensing functions. We derive communication performance metrics of the system in terms of the signal-to-interference-plus-noise ratio (SINR) and communication rate, and derive sensing performance metrics of the system in terms of probability of detection, error of linear least squares (LS) estimation, and the error of linear minimum mean square error (LMMSE) estimation. The proposed joint beamforming framework consists of three stages: tag detection, tag estimation, and communication enhancement. We develop corresponding joint beamforming schemes aimed at enhancing the performance objectives of their respective stages by solving complex non-convex optimization problems. Extensive simulation results demonstrate the effectiveness of the proposed joint beamforming schemes. The proposed B-ISAC system has broad application prospect in next generation IoE scenarios.
\end{abstract}

\IEEEpeerreviewmaketitle

\section{Introduction}
\IEEEPARstart{I}{ntegrated} sensing and communication (ISAC) is a key technology for the next generation wireless communication. ISAC technology aims to integrate sensing and communication, two relatively independent functions in the past, into one system to achieve integration gains and collaboration gains. ISAC plays a crucial role in meeting the diverse sensing and communication requirements of emerging internet of everything (IoE) applications, such as smart homes, smart factories, vehicular networks \cite{Chiriyath2017,Zhang2021,Liu2022,Cui2021,Liu2023,Sturm2011,LiuY2024}. The ISAC system can not only effectively improve spectrum efficiency and hardware efficiency, but also enable the cooperation between sensing and communication to improve the performance of each other.

Researchers from both academia and industry have extensively explored ISAC technology, including system architecture and advanced signal processing techniques \cite{LiuX2020,Tang2022,LiuH2024,Zhou2024,Cheng2024,ZhongK2024,Kwon2023,Wang2023,Kwon2021,Shan2023,LiuF2022}. Moreover, the multiple-input multiple-output (MIMO) architecture is widely adopted in ISAC systems. By leveraging spatial degrees of freedom (DoF), MIMO architecture can provide diversity and multiplexing gains for sensing and communication \cite{LiuX2020}. Therefore, joint waveform/beamforming design becomes a very important problem. Extensive researches on transmit and receive beamforming has been carried out using different communication and sensing metrics\cite{LiuF2018,Yuan2021,Zhao2022,Zhao2024,Cheng2021}. For example, the authors in \cite{LiuF2018} jointly considered the beampattern matching error and multi-user interference (MUI). In \cite{Yuan2021}, the authors jointly optimizes sensing mutual information and communication mutual information. The authors in \cite{Zhao2022} studied the detection performance and proposed a design method to joint optimize the detection probability and communication MUI. Furthermore, the authors in \cite{Zhao2024} proposed a scheme aiming to minimize the maximum Cramér–Rao bound (CRB) of targets under the constraint of signal-to-interference-plus-noise ratio (SINR) for communication. The authors in \cite{Cheng2021} considered the hybrid beamforming problem to maximize the sum-rate under the power constraint and beamformer similarity constraint. In \cite{LiuMM2024}, the authors studied the beamforming problem in reconfigurable intelligent surface (RIS)  assisted ISAC system to achieve the rate maximization under the constraints of power and detection performance.

There is no doubt that these advanced processing schemes can enhance communication and sensing capabilities of conventional ISAC systems. However, conventional ISAC architecture still perceive non-cooperative targets, relying on the echo signal from the target for sensing. The different scattering capabilities of different objects will bring great challenges to the sensing performance. Moreover, the non-cooperative sensing target of the traditional ISAC system are usually considered as silent objectives without providing any information to the system.

The emerging technology of backscatter communication (BackCom) shows significant potential in low-power communications. BackCom uses radio frequency (RF) tags to enable passive communication links by scattering RF signals to the reading device to achieve cognitive spectrum sharing\cite{Kishore2019,Hoang2020,Niu2019,Xu2019,Hu2022}. Researchers have conducted extensive researches on BackCom technology in terms of system architecture\cite{Yang2018,Liang2020,Liang2022,Ren2024}, signal processing\cite{Wan2024,Yang2023,Zhang2024,Xu2023}, multiple access\cite{Li2023,Li2024,Yang2021}, interference cancellation\cite{LiuJ2022}, and so on. These inexpensive tags consume little power while possessing strong signal reflection capabilities. Therefore, the combination of BackCom with ISAC is anticipated to enable energy-efficient passive communication while improving the sensing performance. By placing passive tags with strong scattering capabilities on traditional sensing targets in a ISAC system, the sensing performance of the ISAC system can be significantly enhanced \cite{Decarli2014}. Moreover, the tag allows the target to actively transmit data while being sensed. The integration of BackCom and ISAC may offer a revolutionary solution solution for high-accuracy sensing and low-power communication. In order to address the limitation of conventional ISAC architecture and enable a wider range of applications, we design a backscatter ISAC architecture in this paper, which combines ISAC with BackCom to enable concurrent sensing and communication in IoE networks.
 
Researchers have started exploring the combination of BackCom and ISAC technologies due to the above advantages. The authors in \cite{Gala2023} proposed an integrated sensing
and backscatter communication (ISABC) system and designed a power allocation scheme to optimize its communication rate and sensing rate performance. The authors in \cite{Zarg2023} provided an overview of the sensing and backscatter communication
integration, detailing the development processes and challenges. The authors in \cite{Huang2022} designed a protocol of backscatter ISAC and proposed the processing procedures and algorithms to facilitate joint localization and information transmission. In \cite{LiS2023}, the authors considered a novel ambient backscatter communication-aided ISAC system and proposed a transmission frame structure and a novel estimation algorithm. The authors in \cite{Ren2023} designed a symbiotic localization and the ambient backscatter communication architecture, and developed methods to enable communication and sensing capabilities. The authors in \cite{Venturino2023} exploited the radar clutter as a carrier signal to enable an ambient backscatter communication. In \cite{Tao2024}. authors designed a two-dimensional (2D) direction-of-arrival (DoA) sensing algorithm and a multi-tag symbol detection method. To further consider the detection requirement of RF tags, a joint beamforming design is developed in \cite{Luo2023} for an ISAC system with backscatter RF tags, which could minimize the total transmit power while meeting the tag detection and communication requirements. 

The above works on beamforming design to optimize system performance is still have the following limitations. Existing joint beamforming works consider only sensing rate and detection SINR requirement for sensing performance, while research on estimation performance is still limited. The design of the B-ISAC system in different task modes also needs to be further studied.

\begin{figure*}[!t]
\captionsetup{font=small}
\begin{center}
\includegraphics[width=0.7\textwidth,]{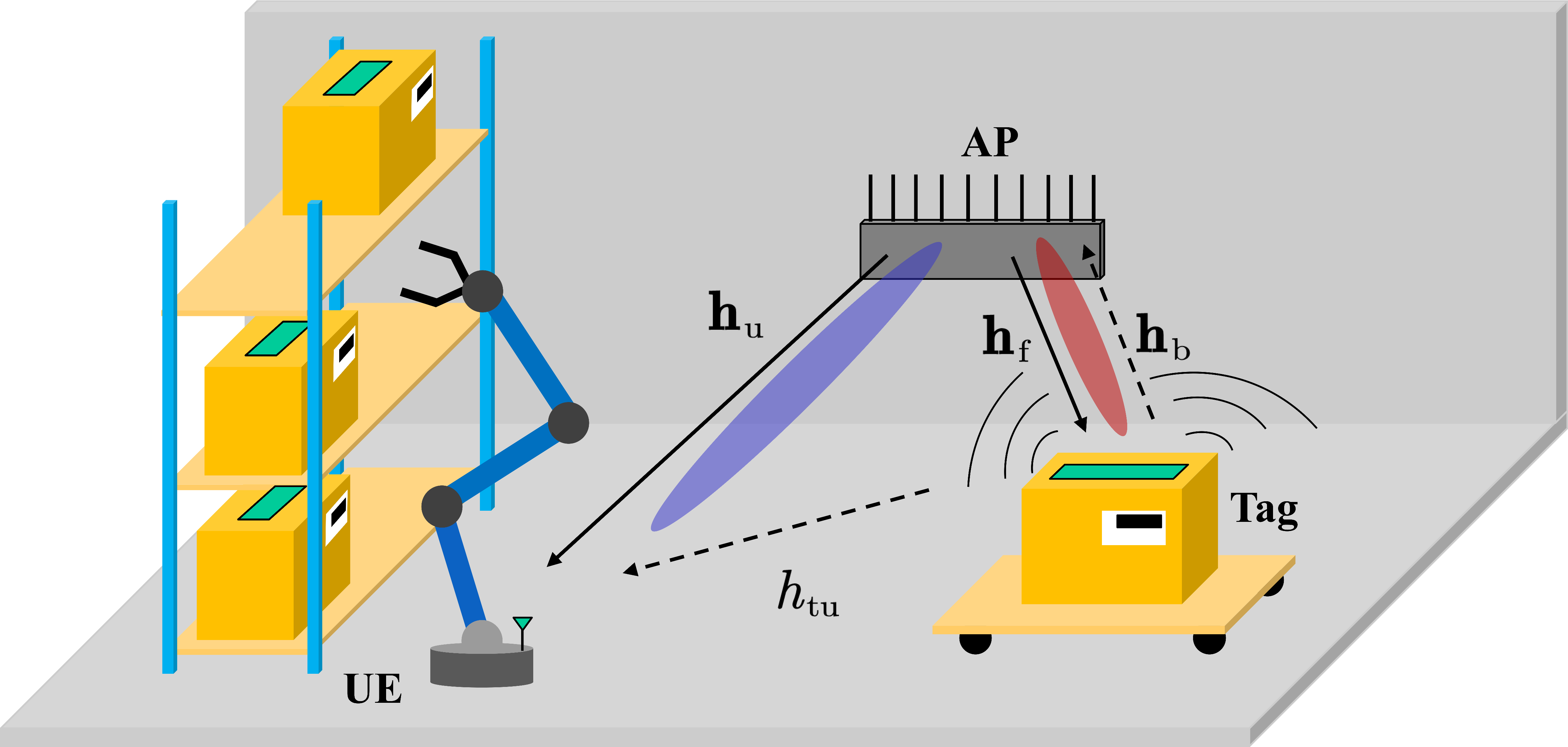}
\end{center}
\vspace{-0.2cm}
\caption{B-ISAC system model for the typical IoE application.}
\label{Fig1}
\end{figure*}

Inspired by the above these prior works, in this paper we focus on designing a backscatter-ISAC (B-ISAC) system, as shown in Fig.~\ref{Fig1}. It aims to exploit a multi-antenna access point (AP) to provide communication service for a user equipment (UE) while also sensing and communicating with a passive RF tag. This system can achieve cognitive spectrum sharing between legacy communication, tag communication and sensing functions. We develop a framework for beamforming design that optimizes the detection, estimation, and communication performance of different stages (task modes) while reducing interference with each other in the B-ISAC system. 

The main contributions of this paper are summarized as follows:
\begin{itemize}
\item First, we formulate the signal model of the B-ISAC system. Specifically, we model the signals received at the tag, UE, and AP. Further, we establish the communication and sensing model and derive the corresponding performance metrics of the B-ISAC system. For communication metric, we derive the SINR and communication rate. For sensing metric, probability of detection, estimation error
of linear least squares (LS) estimation, and the estimation error of linear minimum mean square error (LMMSE) estimation are derived. 

\item Second, we design an optimization framework for joint beamforming of different stages in the B-ISAC system. The proposed framework contains three stages of
tag detection, tag estimation, and communication enhancement. For the tag detection and estimation stages, we formulate three optimization problems to optimize the detection probability, the LS estimation error, and the LMMSE estimation error, respectively, under the constraints of communication SINR performance and power budget. A semidefinite relaxation (SDR) optimal algorithm (Algorithm \ref{alg1}) is developed for solving these three complicated non-convex optimization problems. The proposed SDR optimal algorithm can be used to obtain the global optimal solutions for these problems.

\item Third, for communication enhancement stage, we formulate an optimization problem that maximize the UE communication rate subject to the SINR constraint at both the tag and AP, and total power budget constraint. This formulation ensures that the UE communication rate is maximized when the tag is activated and functioning normally. We develop an iterative algorithm (Algorithm \ref{alg3}) and a successive convex approximation (SCA) algorithm (Algorithm \ref{alg2}) to solve this complicated non-convex problem.

\item Last, we conduct thorough simulation studies to validate the effectiveness of the proposed schemes and their associated algorithms. Numerical results show that the proposed schemes can achieve better performance compared with other schemes, as well as illustrate the trade-off between communication and sensing performance. 
\end{itemize}
The B-ISAC system fully utilizes the passive back scatter properties of RF tags to achieve low-power communication and sensing capabilities. It is expected to has broad application prospect in sixth generation (6G) IoE scenarios.
 
The remainder of this paper is organized as follows. Sec.~\ref{sec2} introduces the B-ISAC system and signal model. Joint Beamforming framework for the B-ISAC system is proposed in Sec.~\ref{sec3}. We present numerical results in Sec.~\ref{sec4}. Finally, the conclusions are drawn in Sec.~\ref{sec5}.

\emph{Notation}: In this paper, boldface lower-case and upper-case letters indicate vectors and matrices respectively. $\mathbb{R}$ and $\mathbb{C}$ represent the real and complex sets respectively. $||\cdot||_{F}$, $||\cdot||$ and $|\cdot|$ are Frobenius norm, Euclidean norm and absolute value respectively. $\left( \cdot \right)^{-1}$, $\left( \cdot \right)^{\dagger}$, $\left( \cdot \right)^T$, $\left( \cdot\right)^*$, and $\left( \cdot \right)^H$ represent the inverse, pseudo inverse, transpose, complex conjugate, and Hermitian transpose, respectively. $\mathbb{E}\left( \cdot \right)$ represents statistical expectation. $\mathrm{Re}\left\{ \cdot\right\}$ returns the real part of a complex number. $j$ is the imaginary unit, which means $j^2=-1$. $\mathbf{I}_N$ is the $N\times N$ identity matrix. $\mathbf{1}=\left[1,1,\ldots,1\right]^T\in\mathbb{R}^N$. $\mathbf{A}\succeq0$ means that $\mathbf{A}$ is a positive semidefinite matrix. $\odot$ represents the Hadamard product. $\mathrm{diag}\left(\mathbf{a} \right)$ returns a diagonal matrix, the vector composed of its diagonal elements is $\mathbf{a}$. $\mathrm{Tr}\left(\mathbf{A} \right)$ and $\mathrm{rank}\left(\mathbf{A}\right)$ compute the trace and rank of matrix $\mathbf{A}$ respectively. $\mathrm{chol}\left(\mathbf{A}\right)$ returns the Cholesky decomposition of matrix $\mathbf{A}$. $\mathrm{vec}\left(\mathbf{A}\right)$ vectorizes matrix $\mathbf{A}$ by column-stacking.

\section{B-ISAC System and Signal Moedl} \label{sec2}
\subsection{System Model}
The B-ISAC system consists of an AP, a UE, and a passive RF tag, as depicted in Fig.~\ref{Fig1}. The AP intends to provide communication service to the UE while simultaneously sensing and communicating with the passive RF tag. The RF tag receives the signal from the AP to obtain the downlink information, and at the same time modulates the uplink information on the back scattered signal to achieve two-way information transmission. Meanwhile, the reflected echo received by the AP can also be further processed to obtain sensing information such as the location and status of the tag to enable the sensing capability of the system. However, the signals reflected by the tag are considered as interference for the UE.

It is assumed that the transmit and receive array of the AP are uniform linear arrays (ULAs) with $N_t$ and $N_r$ antenna elements respectively and $\lambda/2$ inter-spacing between neighboring antenna elements, where $\lambda$ is the carrier wavelength. Without loss of generality, we assume $N_t \leqslant N_r$. The RF tag and UE are equipped with a single antenna, respectively. We also assume that there is no self-interference between the transmit and receive arrays.

Let $\mathbf{X}\in \mathbb{C} ^{{N_t}\times L}$ denote the transmitted signal by the AP, where $L$ is the length of the signal sample in the time domain such that $L>N_t$.  The signal $\mathbf{X}$ can be expressed as
\begin{gather}
\begin{aligned}
\mathbf{X}=\mathbf{WS}=\mathbf{w}_{\mathrm{u}}\mathbf{s}_{\mathrm{u}}^{H}+\mathbf{w}_{\mathrm{t}}\mathbf{s}_{\mathrm{t}}^{H}+\mathbf{W}_{\mathrm{s}}\mathbf{S}_{\mathrm{s}}\in \mathbb{C} ^{N_t\times L},
\label{eq1}
\end{aligned}
\end{gather}
where $\mathbf{W}$ and $\mathbf{S}$ are respectively joint beamforming matrix and data augmentation matrix given by
\begin{align}
\mathbf{W}&=\left[\mathbf{w}_{\mathrm{u}},\mathbf{w}_{\mathrm{t}},\mathbf{w}_{1},\mathbf{w}_{2},\ldots,\mathbf{w}_{N_t} \right] \in \mathbb{C} ^{N_t\times (N_t+2)},\\
\mathbf{S}&=\left[ \mathbf{s}_{\mathrm{u}},\mathbf{s}_{\mathrm{t}},\mathbf{s}_1, \mathbf{s}_2 \ldots,\mathbf{s}_{N_t} \right] ^H\in \mathbb{C} ^{{(N_t+2)}\times L},
\end{align}
where $\mathbf{w}_{\mathrm{u}}\in \mathbb{C} ^{N_t\times {1}}$ and $\mathbf{w}_{\mathrm{t}}\in \mathbb{C} ^{N_t\times {1}}$ are beamforming vectors for the communication user and RF tag, , respectively. Vectors $\mathbf{s}_{\mathrm{u}}\in \mathbb{C} ^{L\times {1}}$ and $\mathbf{s}_{\mathrm{t}}\in \mathbb{C} ^{L\times {1}}$ are data streams for the UE and RF tag, respectively. To compensate the loss of sensing DoF of the transmit waveform, we incorporate an additional probing stream structure $\left[\mathbf{s}_1,\mathbf{s}_2,...,\mathbf{s}_{N_t} \right]^H =\mathbf{S}_{\mathrm{s}}\in \mathbb{C}^{N_t\times L}$ to the transmit waveform. This can extend the DoF of the transmit waveform to its maximum, i.e., $N_t$ \cite{LiuX2020,LiuF2022}. Note that the dedicated probing streams are only used for sensing and do not carry any information. The extra dedicated probing streams have the auxiliary beamfoming matrix of $\left[\mathbf{w}_1,\mathbf{w}_2,...,\mathbf{w}_{N_t} \right] =\mathbf{W}_{\mathrm{s}}\in \mathbb{C}^{N_t\times N_t}$.

The dedicated probing stream $\mathbf{S}_{\mathrm{s}}$ is carefully designed to satisfy $\frac{1}{L}\mathbf{S}_{\mathrm{s}}{\mathbf{S}_{\mathrm{s}}}^H=\mathbf{I}_{N_t}$. When $L$ is sufficiently large, it is assumed that there is no correlation between dedicated probing streams $\mathbf{S}_{\mathrm{s}}$ and random data stream $[\mathbf{s}_{\mathrm{u}},\mathbf{s}_{\mathrm{t}}]^H$. At the same time, as $L$  increases, the correlation between the two random data streams gradually decreases, satisfying $\frac{1}{L}[\mathbf{s}_{\mathrm{u}},\mathbf{s}_{\mathrm{t}}]^H[\mathbf{s}_{\mathrm{u}},\mathbf{s}_{\mathrm{t}}] \approx \mathbf{I}_{2}$. Based on the above, the data augmentation matrix $\mathbf{S}$ satisfies
\begin{gather}
\begin{aligned}
\frac{1}{L}\mathbf{SS}^H\approx\mathbf{I}_{N_t+2}.
\end{aligned}
\end{gather}
Therefore, the sample covariance matrix of waveform $\mathbf{X}$ can be expressed as\footnote {Please note that this approximation is widely adopted in the existing ISAC works \cite{LiuX2020,LiuF2022,LiuF2018}.}
\begin{gather}
\begin{aligned}
\mathbf{R}_\mathbf{X}=\frac{1}{L}\mathbf{XX}^H\approx\mathbf{WW}^H\in \mathbb{C} ^{N_t \times N_t},
\label{eq5}
\end{aligned}
\end{gather}
which implies $\mathbf{R}_\mathbf{X}$ is only related to the beamforming matrix $\mathbf{W}$. Then the beam pattern of the transmit waveform is
\begin{gather}
\begin{aligned}
P\left( \theta \right) = \mathbf{a}^H\left( \theta \right) \mathbf{R}_\mathbf{X}\mathbf{a}\left( \theta \right),
\label{eq6}
\end{aligned}
\end{gather}
where $\mathbf{a}\left( \theta \right)= \left[ 1,e^{j\pi \sin \theta},e^{2j\pi \sin \theta},...,e^{\left( N_t-1 \right) j\pi \sin \theta} \right] ^T \in \mathbb{C} ^{1 \times N_t}$ is the steering vector of the transmit array, $\theta$ is the azimuth angle relative to the array.

\subsection{Backscatter Model}
The RF tag receives the signal to decode the information, and then modulates the information onto the backscattered signal. The signal $\mathbf{y}_{\mathrm{t}}\in \mathbb{C} ^{1\times L}$ received by the RF tag can be expressed as
\begin{align}
\mathbf{y}_{\mathrm{t}}&=\mathbf{h}_{\mathrm{f}}\mathbf{X}+\mathbf{n}_{\mathrm{t}} \nonumber
\\
&=\mathbf{h}_{\mathrm{f}}\mathbf{w}_{\mathrm{u}}\mathbf{s}_{\mathrm{u}}^{H}+\mathbf{h}_{\mathrm{f}}\mathbf{w}_{\mathrm{t}}\mathbf{s}_{\mathrm{t}}^{H}+\mathbf{h}_{\mathrm{f}}\mathbf{W}_{\mathbf{s}}\mathbf{S}_{\mathbf{s}}+\mathbf{n}_{\mathrm{t}},
\label{eq7}
\end{align}
where $\mathbf{h}_{\mathrm{f}}\in \mathbb{C} ^{1\times N_t}$ is the channel vector from the AP to RF tag, $\mathbf{n}_{\mathrm{t}}\in \mathbb{C} ^{1\times L}$ is the receiver noise vector at the tag, which is assumed to follow a zero-mean complex Gaussian distribution as $\mathbf{n}_{\mathrm{t}}\sim \mathcal C\mathcal N \left( 0,\sigma_{\mathrm{t}}^{2}\mathbf{I}_L \right)$. On the right hand side of (\ref{eq7}), the second part $\mathbf{h}_{\mathrm{f}}\mathbf{w}_{\mathrm{t}}\mathbf{s}_{\mathrm{t}}^{H}$ is the desired signal for the tag, the first part $\mathbf{h}_{\mathrm{f}}\mathbf{w}_{\mathrm{u}}\mathbf{s}_{\mathrm{u}}^{H}$ is the interference caused by the legacy communication data, the third part $\mathbf{h}_{\mathrm{f}}\mathbf{W}_{\mathbf{s}}\mathbf{S}_{\mathbf{s}}$ is the interference caused by the dedicated probing stream, and $\mathbf{n}_{\mathrm{t}}$ is the noise vector. Therefore, the SINR of the signal received at the tag can be expressed as
\begin{align}
\gamma _{\mathrm{t}}&=\frac{\mathbb{E} \left( \|\mathbf{h}_{\mathrm{f}}\mathbf{w}_{\mathrm{t}}\mathbf{s}_{\mathrm{t}}^{H}\|^2 \right)}{\mathbb{E} \left( \|\mathbf{h}_{\mathrm{f}}\mathbf{w}_{\mathrm{u}}\mathbf{s}_{\mathrm{u}}^{H}\|^2 \right) +\mathbb{E} \left( \left\| \mathbf{h}_{\mathrm{f}}\mathbf{W}_{\mathrm{s}}\mathbf{S}_{\mathrm{s}} \right\| ^2 \right) +\mathbb{E} \left( \|\mathbf{n}_{\mathrm{t}}\|^2 \right)} \nonumber
\\ 
&=\frac{|\mathbf{h}_{\mathrm{f}}\mathbf{w}_{\mathrm{t}}|^2}{|\mathbf{h}_{\mathrm{f}}\mathbf{w}_{\mathrm{u}}|^2+\left\| \mathbf{h}_{\mathrm{f}}\mathbf{W}_{\mathrm{s}} \right\| ^2+\sigma _{\mathrm{t}}^{2}}.
\label{eq8}
\end{align}
Then, the RF tag encodes the information into the backscatter signal $\mathbf{y}_{\mathrm{b}}\in \mathbb{C} ^{1\times L}$, which is given by
\begin{align}
\mathbf{y}_{\mathrm{b}}=&\sqrt{\alpha}\mathbf{y}_{\mathrm{t}} \odot  \mathbf{c}_{\mathrm{t}}\nonumber
 \\
=&\sqrt{\alpha}\mathbf{h}_{\mathrm{f}}\mathbf{X}\odot \mathbf{c}_{\mathrm{t}}+\sqrt{\alpha}\mathbf{n}_{\mathrm{t}}\odot  \mathbf{c}_{\mathrm{t}}\nonumber
\\
=&\sqrt{\alpha}\mathbf{h}_{\mathrm{f}}\mathbf{w}_{\mathrm{u}}\mathbf{s}_{\mathrm{u}}^{H}\odot \mathbf{c}_{\mathrm{t}}+\sqrt{\alpha}\mathbf{h}_{\mathrm{f}}\mathbf{w}_{\mathrm{t}}\mathbf{s}_{\mathrm{t}}^{H}\odot \mathbf{c}_{\mathrm{t}}\nonumber\\
&+\sum_{i=1}^{N_t}{\sqrt{\alpha}\mathbf{h}_{\mathrm{f}}\mathbf{w}_i\mathbf{s}_i^{H}\odot \mathbf{c}_t}+\sqrt{\alpha}\mathbf{n}_{\mathrm{t}}\odot \mathbf{c}_{\mathrm{t}},
\end{align}
where $\alpha$ is the backscatter modulation efficiency coefficient, and $\mathbf{c}_t$ is the encoded uplink data with $\mathbb{E} [\|\mathbf{c}_t\|^2] =1 $.

We denote the channel from the RF tag to AP as $\mathbf{h}_{\mathrm{b}}\in \mathbb{C} ^{1\times N_r}$, then the backscatter signal received at the AP $\mathbf{Y}_{\mathrm{ap}} \in \mathbb{C} ^{N_r \times L} $ can be expressed as
\begin{align}
\mathbf{Y}_{\mathrm{ap}}=&\mathbf{h}_{\mathrm{b}}\mathbf{y}_{\mathrm{b}}+\mathbf{N}_{\mathrm{ap}}\nonumber
\\
=&\sqrt{\alpha}\mathbf{h}_{\mathrm{b}}\mathbf{h}_{\mathrm{f}}\mathbf{X}\odot \mathbf{c}_{\mathrm{t}}+\sqrt{\alpha}\mathbf{h}_{\mathrm{b}}\mathbf{n}_{\mathrm{t}}\odot \mathbf{c}_{\mathrm{t}}+\mathbf{N}_{\mathrm{ap}}\nonumber
\\
=&\sqrt{\alpha}\mathbf{h}_{\mathrm{b}}\mathbf{h}_{\mathrm{f}}\mathbf{w}_{\mathrm{u}}\mathbf{s}_{\mathrm{u}}^{H}\odot \mathbf{c}_{\mathrm{t}}+\sqrt{\alpha}\mathbf{h}_{\mathrm{b}}\mathbf{h}_{\mathrm{f}}\mathbf{w}_{\mathrm{t}}\mathbf{s}_{\mathrm{t}}^{H}\odot \mathbf{c}_{\mathrm{t}}\nonumber
\\
&+\sum_{i=1}^{N_t}{\sqrt{\alpha}\mathbf{h}_{\mathrm{b}}\mathbf{h}_{\mathrm{f}}\mathbf{w}_i\mathbf{s}_i^{H}\odot \mathbf{c}_t}+\sqrt{\alpha}\mathbf{h}_{\mathrm{b}}\mathbf{n}_{\mathrm{t}}\odot \mathbf{c}_{\mathrm{t}}+\mathbf{N}_{\mathrm{ap}},
\end{align}
where $\mathbf{N}_{\mathrm{ap}}\in \mathbb{C} ^{N_r\times L}$ is the noise matrix at the AP, which is assumed as $\mathrm{vec}(\mathbf{N}_{\mathrm{ap}})\sim \mathcal C\mathcal N \left( 0,\sigma_{\mathrm{ap}}^{2}\mathbf{I}_{{N_r}L} \right)$.
The combined signal $\tilde{\mathbf{y}}_{\mathrm{ap}} \in \mathbb{C} ^{1\times L}$ is given by
\begin{align}
\tilde{\mathbf{y}}_{\mathrm{ap}}=&\mathbf{w}_{\mathrm{r}}\mathbf{Y}_{\mathrm{ap}}\nonumber
\\
=&\sqrt{\alpha}\mathbf{w}_{\mathrm{r}}\mathbf{h}_{\mathrm{b}}\mathbf{h}_{\mathrm{f}}\mathbf{X}\odot \mathbf{c}_{\mathrm{t}}+\sqrt{\alpha}\mathbf{w}_{\mathrm{r}}\mathbf{h}_{\mathrm{b}}\mathbf{n}_{\mathrm{t}}\odot \mathbf{c}_{\mathrm{t}}+\mathbf{w}_{\mathrm{r}}\mathbf{N}_{\mathrm{ap}}.\nonumber
\\
=&\sqrt{\alpha}\mathbf{w}_{\mathrm{r}}\mathbf{h}_{\mathrm{b}}\mathbf{h}_{\mathrm{f}}\mathbf{w}_{\mathrm{u}}\mathbf{s}_{\mathrm{u}}^{H}\odot  \mathbf{c}_{\mathrm{t}}+\sqrt{\alpha}\mathbf{w}_{\mathrm{r}}\mathbf{h}_{\mathrm{b}}\mathbf{h}_{\mathrm{f}}\mathbf{w}_{\mathrm{t}}\mathbf{s}_{\mathrm{t}}^{H}\odot \mathbf{c}_{\mathrm{t}}\nonumber
\\
&+\sum_{i=1}^{N_t}{\sqrt{\alpha}\mathbf{w}_{\mathrm{r}}\mathbf{h}_{\mathrm{b}}\mathbf{h}_{\mathrm{f}}\mathbf{w}_i\mathbf{s}_i^{H}\odot\mathbf{c}_t}+\sqrt{\alpha}\mathbf{w}_{\mathrm{r}}\mathbf{h}_{\mathrm{b}}\mathbf{n}_{\mathrm{t}}\odot \mathbf{c}_{\mathrm{t}}\nonumber\\
&+\mathbf{w}_{\mathrm{r}}\mathbf{N}_{\mathrm{ap}}
\label{eq11}
\end{align}
where $\mathbf{w}_{\mathrm{r}}\in \mathbb{C} ^{1\times N_r}$ is the combining vector. It is worth noting that the transmit signal $\mathbf{X}$ is completely known to the AP, so $\mathbf{X}$ can be used to perform matched filtering and decoding the information. We denote $\widetilde{\mathbf{C}}_{\mathrm{t}}=\mathrm{diag}\left( \mathbf{c}_{\mathrm{t}} \right)$. Consequently, the received SINR at the AP is 
\begin{align}
\gamma _{\mathrm{ap}}&=\frac{\mathbb{E} \left( \left\| \sqrt{\alpha}\mathbf{w}_{\mathrm{r}}\mathbf{h}_{\mathrm{b}}\mathbf{h}_{\mathrm{f}}\mathbf{X}\odot \mathbf{c}_{\mathrm{t}} \right\| ^2 \right)}{\mathbb{E} \left( \left\| \sqrt{\alpha}\mathbf{w}_{\mathrm{r}}\mathbf{h}_{\mathrm{b}}\mathbf{n}_{\mathrm{t}}\odot \mathbf{c}_{\mathrm{t}} \right\| ^2 \right) +\mathbb{E} \left( \left\| \mathbf{w}_{\mathrm{r}}\mathbf{N}_{\mathrm{AP}} \right\| ^2 \right)}\nonumber
\\
&=\frac{\alpha \mathbb{E} \left( \mathbf{w}_{\mathrm{r}}\mathbf{h}_{\mathrm{b}}\mathbf{h}_{\mathrm{f}}\mathbf{X}\widetilde{\mathbf{C}}_{\mathrm{t}} \widetilde{\mathbf{C}}_{\mathrm{t}} ^H\mathbf{X}^H\mathbf{h}_{\mathrm{f}}^{H}\mathbf{h}_{\mathrm{b}}^{H}\mathbf{w}_{\mathrm{r}}^{H} \right)}{\alpha \mathbb{E} \left( |\mathbf{w}_{\mathrm{r}}\mathbf{h}_{\mathrm{b}}\mathbf{n}_{\mathrm{t}}\widetilde{\mathbf{C}}_{\mathrm{t}} \widetilde{\mathbf{C}}_{\mathrm{t}}^H\mathbf{n}_{\mathrm{t}}^{H}\mathbf{h}_{\mathrm{b}}^{H}\mathbf{w}_{\mathrm{r}}^{H}| \right) +\left\| \mathbf{w}_{\mathrm{r}} \right\| ^2\sigma _{\mathrm{ap}}^{2}}\nonumber
\\
&\overset{\left( a \right)}{=}\frac{\alpha  \mathbf{w}_{\mathrm{r}}\mathbf{h}_{\mathrm{b}}\mathbf{h}_{\mathrm{f}}\mathbf{R}_{\mathbf{X}}\mathbf{h}_{\mathrm{f}}^{H}\mathbf{h}_{\mathrm{b}}^{H}\mathbf{w}_{\mathrm{r}}^{H}}{\alpha |\mathbf{w}_{\mathrm{r}}\mathbf{h}_{\mathrm{b}}|^2\sigma _{\mathrm{t}}^{2}+\left\| \mathbf{w}_{\mathrm{r}} \right\| ^2\sigma _{\mathrm{ap}}^{2}},
\label{eq12}
\end{align}
where ($a$) holds because $\mathbb{E} \left(\widetilde{\mathbf{C}}_{\mathrm{t}} \widetilde{\mathbf{C}}_{\mathrm{t}}^H \right)={\mathbf{I}}_{L}$.
For convenience, when the channel $\mathbf{h}_{\mathrm{b}}$ is known to the AP, we use equal gain combining vector, i.e., $\mathbf{w}_{\mathrm{r}}=\mathbf{h}_{\mathrm{b}}/\left\| \mathbf{h}_{\mathrm{b}} \right\|$. While when $\mathbf{h}_{\mathrm{b}}$ is unknown, detailed explanation are given in the sub-section \ref{2C}.
\begin{Remark}
Since $\mathbf{X}$ is completely known to the AP, the SINR of the received signal at the AP is related to the sample covariance matrix of $\mathbf{X}$, not just related to the signal $\mathbf{w}_{\mathrm{t}}\mathbf{s}_{\mathrm{t}}$, which is different from \cite{Luo2023}. In other words, communication signal $\mathbf{w}_{\mathrm{u}}\mathbf{s}_{\mathrm{u}}$ and dedicated probing stream $\mathbf{W}_{\mathrm{s}}\mathbf{S}_{\mathrm{s}}$  can also help the tag to facilitate backscatter communication.
\end{Remark}
\begin{Remark}
The data stream $\mathbf{s}_{\mathrm{t}}$ contains a sequence that can activate the passive RF tag for data transmission. Therefore, in order to detect the tag and complete communication transmission, the SINRs at the tag and AP must be greater than a certain threshold to meet their respective sensitivity constraints.
\end{Remark}

\subsection{Sensing Model}
\label{2C}
\textbf{Detection}: For the sensing model of the B-ISAC system, the first task is the tag detection, aiming to determine whether the RF tag is present in the environment. We can use the signal received by the AP to perform a signal detection process. Note that when the AP is on sensing mode, we assume that the tag is not activated, which means $\mathbf{c}_\mathrm{t}=\mathbf{1}_{L}$.
We can formulate the detection promblem as a hypothesis testing problem as follows,
\begin{align}
\begin{cases}
	\mathcal{H} _0:\tilde{\mathbf{y}}_{\mathrm{ap}}=\mathbf{w}_{\mathrm{r}}\mathbf{N}_{\mathrm{AP}}\\
	\mathcal{H} _1:\tilde{\mathbf{y}}_{\mathrm{ap}}=[\sqrt{\alpha}\mathbf{w}_{\mathrm{r}}\mathbf{h}_{\mathrm{b}}\mathbf{h}_{\mathrm{f}}(\mathbf{X}+\mathbf{n}_{\mathrm{t}})]\odot \mathbf{1}_{L}+\mathbf{w}_{\mathrm{r}}\mathbf{N}_{\mathrm{ap}},
\end{cases}
\label{eq10}
\end{align}
where $\mathcal{H} _0$ means there is no backscatter signal from the tag, $\mathcal{H} _1$ means there is tag echo.
According to Neyman-Pearson criterion \cite{Neyman1992,Kay1998,Fortu2020,Tang2022}, we can get the following detector
\begin{gather}
\begin{aligned}
\mathrm{Re}\left\{ \tilde{\mathbf{y}}_{\mathrm{ap}} (\sqrt{\alpha}\mathbf{w}_{\mathrm{r}}\mathbf{h}_{\mathrm{b}}\mathbf{h}_{\mathrm{f}}\mathbf{X})^H \right\} 
 \begin{array}{c}
	\overset{\mathcal{H}_1}{\geqslant}\\
	\underset{\mathcal{H}_0}{<}\\
\end{array}\,\, \eta,
 \label{eq14}
\end{aligned}
\end{gather}
where $\eta$ is the detection threshold. The detection probability of the tag denoted as $P_D$ is given by \cite{Tang2022}
\begin{gather}
\begin{aligned}
P_D=\frac{1}{2}\mathrm{erfc}\left\{ \mathrm{erfc}^{-1}\left( 2P_{F} \right) -\sqrt{\gamma _{\mathrm{ap}}} \right\},
\label{eq15}
\end{aligned}
\end{gather}
where $\mathrm{erfc}\left( x \right) =\frac{2}{\sqrt{\pi}}\int_x^{\infty}{e^{-t^2}}dt$ is  the complementary error function. $P_{F}$ is the probability of false alarm and is set to be a constant when using constant false-alarm rate (CFAR) detection.

For the detection problem, whether the tag exists or not is usually unknown to the AP, therefore it is hard to obtain the channel $\mathbf{h}_{\mathrm{b}}$ and $\mathbf{h}_{\mathrm{f}}$. To address this issue, we divide the continuous angle space into several discrete grids, and detect each angle grid separately. Then, we use the receiving steering vector $\mathbf{b}\left( \theta \right)= \left[ 1,e^{j\pi \sin \theta},e^{2j\pi \sin \theta},...,e^{\left( N_r-1 \right) j\pi \sin \theta} \right] ^T \in \mathbb{C} ^{1 \times N_r}$ to formulate the combining vector as $\mathbf{b}\left( \theta \right)/\left\| \mathbf{b}\left( \theta \right) \right\|$ for effective detection. 

In addition to the detection task, estimation of the tag position is also very essential. To formulate the estimation task, we define $\mathbf{G}=\mathbf{h}_{\mathrm{b}}\mathbf{h}_{\mathrm{f}}$, $\tilde{\mathbf{X}}=\mathbf{X}\widetilde{\mathbf{C}}_{\mathrm{t}} $, and $\sqrt{\alpha}\mathbf{h}_{\mathrm{b}}\mathbf{n}_{\mathrm{t}}\odot \mathbf{c}_{\mathrm{t}}+\mathbf{N}_{\mathrm{ap}}=\tilde{\mathbf{N}}_{\mathrm{ap}}$, where $\mathrm{vec}(\tilde{\mathbf{N}}_{\mathrm{ap}})\sim \mathcal C\mathcal N \left( 0,\sigma _{\mathrm{ap}}^{2}+\alpha \left\| \mathbf{h}_{\mathrm{b}} \right\| ^2\sigma _{\mathrm{t}}^{2}\mathbf{I}_{{N_r}L} \right)$. According to (\ref{eq11}), the received signal at the AP is rewritten as
\begin{gather}
\begin{aligned}
\mathbf{Y}_{\mathrm{ap}}=\sqrt{\alpha}\mathbf{G}\tilde{\mathbf{X}}+\tilde{\mathbf{N}}_{\mathrm{ap}}.
\label{eq16}
\end{aligned}
\end{gather}
All information about the tag position is actually coupled in $\mathbf{G}$. The estimation of the tag position can be regarded as an estimation of $\mathbf{G}$. Therefore, the tag position estimation problem can be regarded as using the ``pilot $\tilde{\mathbf{X}}$''  to estimate the ``channel $\mathbf{G}$''. Note that in the estimation stage, we still assume that $\mathbf{c}_\mathrm{t}=\mathbf{1}_{L}$, i.e., $\tilde{\mathbf{X}}={\mathbf{X}}$.

\textbf{Non-Bayesian Estimation}: Without any prior information about $\mathbf{G}$, we can use the linear least squares (LS) estimator to estimate $\mathbf{G}$. Then, the LS estimator can be expressed as
\begin{gather}
\begin{aligned}
\hat{\mathbf{G}}_{\mathrm{LS}}=\frac{1}{\sqrt{\alpha}}\mathbf{Y}_{\mathrm{ap}}\tilde{\mathbf{X}}^{\dagger},
\end{aligned}
\end{gather}
where $\tilde{\mathbf{X}}^{\dagger}=\tilde{\mathbf{X}}^{H}
(\tilde{\mathbf{X}}\tilde{\mathbf{X}}^{H})^{-1}$ is the pseudo inverse of $\tilde{\mathbf{X}}$. The corresponding estimation error is given by
\begin{align}
\label{eq18}
J_{\mathrm{LS}}=&\mathbb{E}\left\{\left\| \mathbf{G}- \hat{\mathbf{G}}_{\mathrm{LS}}\right\|_{F}^{2}\right\}\nonumber
\\
=&\mathrm{Tr}\left\{ \left( \mathbf{G}-\hat{\mathbf{G}}_{\mathrm{LS}} \right) \left( \mathbf{G}-\hat{\mathbf{G}}_{\mathrm{LS}} \right) ^H \right\} \nonumber
\\
=&\frac{N_r\left( \sigma _{\mathrm{ap}}^{2}+\alpha \left\| \mathbf{h}_{\mathrm{b}} \right\| ^2\sigma _{\mathrm{t}}^{2} \right)}{\alpha }\mathrm{Tr}\left[ \left( \tilde{\mathbf{X}}\tilde{\mathbf{X}}^{H} \right) ^{-1} \right]\nonumber \\
\overset{(b)}{\approx}&\frac{N_r\left( \sigma _{\mathrm{ap}}^{2}+\alpha \left\| \mathbf{h}_{\mathrm{b}} \right\| ^2\sigma _{\mathrm{t}}^{2} \right)}{\alpha L}\mathrm{Tr}\left[ \left( {\mathbf{W}}{\mathbf{W}}^H \right) ^{-1} \right]. 
\end{align}
According to \eqref{eq5}, $\mathbf{XX}^H\approx L\mathbf{WW}^H$. Therefore, the $(b)$ holds. 

\textbf{(Optimal Waveform for LS : Orthogonal) } Any waveform matrix $\mathbf{\hat{X}}_\mathrm{LS}$ under power constraint $P_T$ is optimal for LS estimation if it satisfies the following equation \cite{Biguesh2006}
\begin{gather}
\begin{aligned}
\frac{1}{L}\mathbf{\hat{X}}_\mathrm{LS}\mathbf{\hat {X}}_\mathrm{LS}^H=\frac{P_T}{N_t}\mathbf{I}_{N_t}.  
\label{eq19}
\end{aligned}
\end{gather}
With only energy constraint, the optimal LS estimation waveform is the fully orthogonal waveform. Since there is no prior information, it is optimal to radiate the energy uniformly to the entire space.

\textbf{Bayesian Estimation}: If some prior information about $\mathbf{G}$ is available, such as the matrix of channel correlations $\mathbf{R}_{\mathbf{G}}=\mathbb{E} \left\{ \mathbf{G}^H\mathbf{G} \right\}$,\footnote {Without loss of generality, it is assumed that $\mathbf{R}_{\mathbf{G}}$ is a matrix of full rank, i.e.,  $\mathrm{rank}\left( \mathbf{R}_{\mathbf{G}} \right) =N_t$, consistent with that in \cite{Biguesh2006}.} we can use the linear minimum mean square error (LMMSE) estimator to estimate the $\mathbf{G}$, where the LMMSE estimator can be expressed as equation (\ref{eq20}).
\begin{figure*}
\begin{gather}
\begin{aligned}
&\hat{\mathbf{G}}_{\mathbf{LMMSE}}
=\frac{1}{\sqrt{\mathbf{\alpha }}}\mathbf{Y}_{\mathrm{ap}}\left( \tilde{\mathbf{X}}^H\mathbf{R}_{\mathbf{G}}\tilde{\mathbf{X}}+\frac{N_r\left( \sigma _{\mathrm{ap}}^{2}+\alpha \left\| \mathbf{h}_{\mathrm{b}} \right\| ^2\sigma _{\mathrm{t}}^{2} \right)}{\alpha}\mathbf{I}_L \right) ^{-1}\tilde{\mathbf{X}}^H\mathbf{R}_{\mathbf{G}}.
\label{eq20}
\end{aligned}
\end{gather}
\vspace{-0.5cm}
\end{figure*}

The estimation error of the LMMSE estimator is given by
\begin{align}
\label{eq21}
&J_{\mathrm{LMMSE}}=\mathbb{E}\left\{\left\| \mathbf{G}- \hat{\mathbf{G}}_{\mathrm{LMMSE}}\right\|_{F}^{2}\right\}\nonumber\\
&=\mathrm{Tr}\left\{ \left( \mathbf{G}-\hat{\mathbf{G}}_{\mathrm{LMMSE}} \right) \left( \mathbf{G}-\hat{\mathbf{G}}_{\mathrm{LMMSE}} \right) ^H \right\} \nonumber
\\
&=\mathrm{Tr}\left\{ \left( \mathbf{R}_{\mathbf{G}}^{-1}+\frac{\alpha}{N_r\left( \sigma _{\mathrm{ap}}^{2}+\alpha \left\| \mathbf{h}_{\mathrm{b}} \right\| ^2\sigma _{\mathrm{t}}^{2} \right)}\tilde{\mathbf{X}}\tilde{\mathbf{X}}^H \right) ^{-1} \right\}\nonumber\\ 
&\overset{(c)}{\approx}
 \mathrm{Tr}\left\{ \left( \mathbf{R}_{\mathbf{G}}^{-1}+\frac{\alpha L}{N_r\left( \sigma _{\mathrm{ap}}^{2}+\alpha \left\| \mathbf{h}_{\mathrm{b}} \right\| ^2\sigma _{\mathrm{t}}^{2} \right)}{\mathbf{W}}{\mathbf{W}}^H \right) ^{-1} \right\}.
\end{align}
According to \eqref{eq5}, $\mathbf{XX}^H\approx L\mathbf{WW}^H$. Therefore, the $(c)$ holds.

\begin{Remark}
$\mathbf{R}_{\mathbf{G}}$ is the prior information about ${\mathbf{G}}$. When all eigenvalues of $\mathbf{R}_{\mathbf{G}}$ tend to infinity (the variance of G tends to infinity), the $\mathbf{R}_{\mathbf{G}}^{-1}$ will tend to the zero matrix (no prior information). In this case, the Bayesian estimation will degenerate to the non-Bayesian estimation, and (\ref{eq21}) will degenerate to (\ref{eq18}).
\end{Remark}
\textbf{(Optimal waveform for LMMSE: Water filling) } Optimal waveform $\mathbf{\hat{X}}_\mathrm{LMMSE}$ for LMMSE under power consraint $P_t$ is given by \cite{Biguesh2006}
\begin{gather}
\begin{aligned}
\mathbf{\hat{X}}_\mathrm{LMMSE}=
\beta\mathbf{Q}\left[ \left( \mu _0\mathbf{I}_{N_t}-\mathbf{\Lambda }^{-1} \right) ^+,\mathbf{0}_{N_t\times \left( L-N_t \right)} \right] ^{\frac{1}{2}}\mathbf{U},
\label{eq22}
\end{aligned}
\end{gather}
where $\beta=\sqrt{\frac{N_r\left( \sigma _{\mathrm{ap}}^{2}+\alpha \left\| \mathbf{h}_{\mathrm{b}} \right\| ^2\sigma _{\mathrm{t}}^{2} \right)}{\alpha}}$, $\mathbf{Q}$ and $\mathbf{\Lambda }$ are obtained from eigenvalue decomposition (EVD) of $\mathbf{R}_{\mathbf{G}}$, i.e., $\mathbf{R}_{\mathbf{G}}=\mathbf{Q}\mathbf{\Lambda }\mathbf{Q}^{H}$,  $\mathbf{U} \in \mathbb{C} ^{N_t \times N_t}$ is an arbitrary unitary matrix. $\mu _0$ is a constant which is known as water level to ensure the waveform to meet power constraint.

Therefore, (\ref{eq21}) and (\ref{eq18}) can be used to evaluate the estimation performance of the transmit waveform for Bayesian and non-Bayesian cases, when $L$ is sufficiently large. It means that, in this paper, we could use (\ref{eq21}) and (\ref{eq18}) to approximately evaluate the tag positioning ability of the beamformer Bayesian estimation and non-Bayesian estimation cases, respectively.

\subsection{Communication Model}
In addition to tag sensing and communication, the downlink communication connections with legacy communication UE also needs to be established for the AP of the B-ISAC system. We denote the channel between the AP and UE as $\mathbf{h}_{\mathrm{u}}\in \mathbb{C} ^{1\times N_t}$, and the channel between the tag and UE as $h_{\mathrm{tu}}\in \mathbb{C}$. Then, the received communication signal at the UE $\mathbf{y}_{\mathrm{u}}\in \mathbb{C} ^{1\times L}$ can be expressed as
\begin{align}
\mathbf{y}_{\mathrm{u}}=&\mathbf{h}_{\mathrm{u}}\mathbf{X}+h_{\mathrm{tu}}\mathbf{y}_{\mathrm{b}}+\mathbf{n}_{\mathrm{u}}\nonumber
\\
=&\mathbf{h}_{\mathrm{u}}\mathbf{X}+h_\mathrm{tu}\left( \sqrt{\alpha}\mathbf{h}_\mathrm{u}\mathbf{X}\odot  \mathbf{c}_{\mathrm{t}}+\sqrt{\alpha}\mathbf{n}_{\mathrm{t}}\odot  \mathbf{c}_{\mathrm{t}} \right) +\mathbf{n}_{\mathrm{u}}\nonumber
\\
=&\mathbf{h}_{\mathrm{u}}\mathbf{w}_{\mathrm{u}}\mathbf{s}_{\mathrm{u}}^{H}+\mathbf{h}_{\mathrm{u}}\mathbf{w}_{\mathrm{t}}\mathbf{s}_{\mathrm{t}}^{H}+\sum_{i=1}^{N_t}{\mathbf{h}_{\mathrm{u}}\mathbf{w}_i\mathbf{s}_i^{H}}\nonumber
\\&+h_{\mathrm{tu}}\sqrt{\alpha}\mathbf{h}_{\mathrm{f}}\mathbf{w}_{\mathrm{u}}\mathbf{s}_{\mathrm{u}}^{H}\odot  \mathbf{c}_{\mathrm{t}}+h_{\mathrm{tu}}\sqrt{\alpha}\mathbf{h}_{\mathrm{f}}\mathbf{w}_{\mathrm{t}}\mathbf{s}_{\mathrm{t}}^{H}\odot  \mathbf{c}_{\mathrm{t}}\nonumber
\\&+\sum_{i=1}^{N_t}{h_{\mathrm{tu}}\sqrt{\alpha}\mathbf{h}_{\mathrm{f}}\mathbf{w}_i\mathbf{s}_i^{H}\odot \mathbf{c}_t}+h_{\mathrm{tu}}\sqrt{\alpha}\mathbf{n}_{\mathrm{t}}\odot  \mathbf{c}_{\mathrm{t}}+\mathbf{n}_{\mathrm{u}}.
\label{eq23}
\end{align}
There are three main parts in the right hand side of signal $\mathbf{y}_{\mathrm{u}}$. Signal $\mathbf{h}_{\mathrm{u}}\mathbf{X}$ is from the AP, interference $h_{\mathrm{tu}}\mathbf{y}_{\mathrm{b}}$ is from RF tag, and $\mathbf{n}_{\mathrm{u}}$ is the noise vector at the UE, which is assumed as  $\mathbf{n}_{\mathrm{u}}\sim \mathcal C\mathcal N \left( 0,\sigma_{\mathrm{u}}^{2}\mathbf{I}_L \right)$. The SINR of the signal received by UE is given in (\ref{eq24}).
\begin{figure*}
\begin{align}
\gamma _\mathrm{u}&=\frac{\mathbb{E} \left( \left\| \mathbf{h}_{\mathrm{u}}\mathbf{w}_{\mathrm{u}}\mathbf{s}_{\mathrm{u}}^{H} \right\| ^2 \right)}{\mathbb{E} \left( \left\| \mathbf{h}_{\mathrm{u}}\mathbf{w}_{\mathrm{t}}\mathbf{s}_{\mathrm{t}}^{H} \right\| ^2 \right) +\mathbb{E} \left( \left\| \sum_{i=1}^{N_t}{\mathbf{h}_{\mathrm{u}}\mathbf{w}_i\mathbf{s}_i^{H}} \right\| ^2 \right) +\mathbb{E} \left( \left\| h_\mathrm{tu}\left[ \sqrt{\alpha}\mathbf{h}_{\mathrm{f}}\left( \mathbf{w}_{\mathrm{u}}\mathbf{s}_{\mathrm{u}}^{H}+\mathbf{w}_{\mathrm{t}}\mathbf{s}_{\mathrm{t}}^{H}+\mathbf{W}_{\mathrm{s}}\mathbf{S}_{\mathrm{s}} +\sigma _{\mathrm{t}}^{2} \right) \odot \mathbf{c}_{\mathrm{t}} \right] \right\| ^2 \right) +\mathbb{E} \left( \left\| \mathbf{n}_{\mathrm{u}} \right\| ^2 \right)}\nonumber
\\
&=\frac{|\mathbf{h}_{\mathrm{u}}\mathbf{w}_{\mathrm{u}}|^2}{|\mathbf{h}_{\mathrm{u}}\mathbf{w}_{\mathrm{t}}|^2+\sum_{i=1}^{N_t}{|\mathbf{h}_{\mathrm{u}}\mathbf{w}_i|^2}+\alpha |h_\mathrm{tu}|^2\left( |\mathbf{h}_{\mathrm{f}}\mathbf{w}_{\mathrm{u}}|^2+|\mathbf{h}_{\mathrm{f}}\mathbf{w}_{\mathrm{t}}|^2+\sum_{i=1}^{N_t}{|\mathbf{h}_{\mathrm{f}}\mathbf{w}_i|^2}+\sigma _{\mathrm{t}}^{2} \right) +\sigma _{\mathrm{u}}^{2}}.
\label{eq24}
\end{align}
\hrulefill
\vspace{-0.2cm}
\end{figure*}
Then, the communication rate of UE can be expressed as
\begin{gather}
\begin{aligned}
R={\log _2\left( 1+\gamma _\mathrm{u} \right)}
\end{aligned}
\end{gather}

\section{Joint Beamforming  Framework for B-ISAC Systems}\label{sec3}
In this section, we propose an optimization framework for joint beamforming
in different stages (task modes) of B-ISAC. Specifically, we formulate four optimization problems for different stages of B-ISAC systems to solve the beamforming matrix, and correspondingly we develop three efficient algorithms to solve these complex non-convex optimization problems. 

\begin{figure*}[!t]
\captionsetup{font=small}
\begin{center}
\includegraphics[width=0.75\textwidth, ]{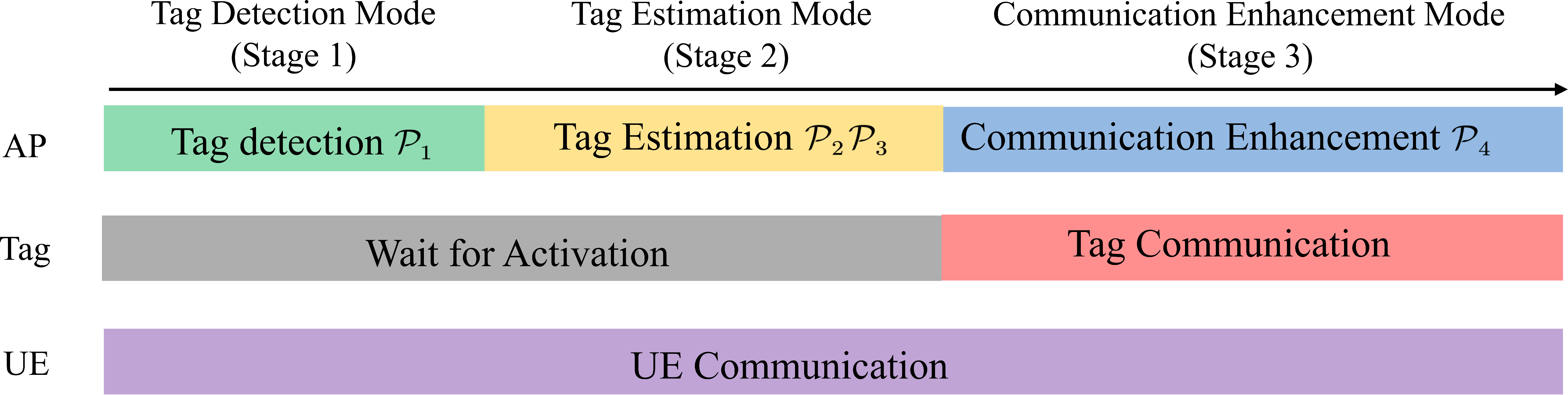}
\end{center}
\caption{Illustration of different stages (task modes) in B-ISAC systems}
\label{Fig22}
\vspace{-0.3cm}
\end{figure*}

We develop three stages for B-ISAC systems: tag detection, tag estimation, and communication enhancement. Stages of B-ISAC systems are shown in Fig.~\ref{Fig22}. The details are as follows,
\begin{itemize}
\item \textbf{\em (Tag detection, stage 1)} It is unknown whether there is tag in the scene. At this time, the AP needs to perform the detection task while communicating with the UE to determine whether there is a tag. We formulate optimization problem $\left( \mathcal{P} _1 \right)$ for this stage.
\item \textbf{\em (Tag estimation, stage 2)} After determining that there is a tag in the scene, the AP needs to further determine the location of the tag and the corresponding communication channel, and ensure communication with the UE. We formulate optimization problem $\left( \mathcal{P} _2 \right)$ and $\left( \mathcal{P} _3 \right)$ for non-Baysian and Bayesian estimation of this stage.
\item \textbf{\em  (Communication enhancement, stage 3)} When the stages 1 and 2 are completely determined, the communication rate of the UE needs to be further improved while ensuring that the tag is activated and can work normally. We formulate optimization problem $\left( \mathcal{P} _4 \right)$ for this stage.
\end{itemize}

From Fig.~\ref{Fig22},it can be seen that, in each task mode, communication and sensing functions are carried out simultaneously. Next, we introduce the joint beamforming scheme in each stage. 

\subsection{Joint Beamforming Scheme for Tag Detection}\label{3A}
Firstly, we consider the situation where the existence of the tag is unknown. In this stage, the B-ISAC system should work in tag detection mode. We consider the design of the beamforming matrix to maximize the probability of detecting the tag. Since the system does not know whether a tag exists, the AP uses spatial scanning to detect targets. Therefore, we formulate the beamforming design problem as an optimization problem that maximizes the detection probability at a certain angle under the constraints of communication SINR and power budget. The problem is given by
\begin{subequations}\label{eq25}
 \begin{align}
\left( \mathcal{P} _1 \right)~~  & \mathop{\mathrm{maximize}} \limits_{\,\,\mathbf{W}}   &&P_D\left( \theta_i \right) 
\\
&~\mathrm{subject~to}   &&\gamma _{\mathrm{u}}\geqslant \gamma _{{\mathrm{uth}}}
\\
& &&\mathrm{Tr}\left( \mathbf{WW}^H \right) \leqslant P_T,
\end{align}
\end{subequations}
where the joint beamforming matrix $\mathbf{W}$ is the optimizing variable. $\theta_i$ is the current angle grid to be detected. $P_D\left( \theta_i \right)$ is the probability of detection, if the tag is at angle $\theta_i$. $P_t$ is the total transmit power, $\mathrm{Tr}\left( \mathbf{WW}^H \right) \leqslant P_t$ is a total power budget constraint for joint beamforming matrix. $\gamma _{\mathrm{u}}\geqslant \gamma _{{\mathrm{uth}}}$ is to set a threshold $\gamma _{{\mathrm{uth}}}$ for the communication SINR of the UE to ensure the communication performance. 

Directly optimizing $P_D(\theta_i)$ is difficult. However, it can be seen from (\ref{eq15}) that maximizing the detection probability is essentially equivalent to maximizing the received signal SINR $\gamma_\mathrm{ap}(\theta_i)$, since the $P_D(\theta_i)$ is a monotonically increasing function of $\gamma_\mathrm{ap}(\theta_i)$. When the tag is at angle $\theta_i$, $\gamma_\mathrm{ap}(\theta_i)$ is the received signal SINR, and the channels $\mathbf{h}_\mathrm{f}$ and $\mathbf{h}_\mathrm{b}$ can be simply regarded as $\mathbf{a}(\theta_i)$ and $\mathbf{b}(\theta_i)^{H}$. Therefore, according to (\ref{eq12}), $\gamma_\mathrm{ap}(\theta_i)$ is given by
\begin{align}
\gamma_\mathrm{ap}(\theta_i)&=
\frac{\alpha \frac{\mathbf{b}\left( \theta _i \right)}{\left\| \mathbf{b}\left( \theta _i \right) \right\|}\mathbf{b}\left( \theta _i \right) ^H\mathbf{a}\left( \theta _i \right) \mathbf{R}_{\mathbf{X}}\mathbf{a}\left( \theta _i \right) ^H\mathbf{b}\left( \theta _i \right) \frac{\mathbf{b}\left( \theta _i \right)}{\left\| \mathbf{b}\left( \theta _i \right) \right\|}^H}{\alpha |\frac{\mathbf{b}\left( \theta _i \right)}{\left\| \mathbf{b}\left( \theta _i \right) \right\|}\mathbf{b}\left( \theta _i \right) ^H|^2\sigma _{\mathrm{t}}^{2}+\left\| \frac{\mathbf{b}\left( \theta _i \right)}{\left\| \mathbf{b}\left( \theta _i \right) \right\|} \right\| ^2\sigma _{\mathrm{ap}}^{2}}\nonumber\\
&=\frac{\alpha N_r\mathbf{a}\left( \theta _i \right) \mathbf{R}_{\mathbf{X}}\mathbf{a}\left( \theta _i \right) ^H}{\alpha N_r\sigma _{\mathrm{t}}^{2}+\sigma _{\mathrm{ap}}^{2}}\nonumber
\\
&\overset{(d)}{\approx} \frac{\alpha N_r\mathbf{a}\left( \theta _i \right) \mathbf{WW}^H\mathbf{a}\left( \theta _i \right) ^H}{\alpha N_r\sigma _{\mathrm{t}}^{2}+\sigma _{\mathrm{ap}}^{2}}.
\end{align}
According to \eqref{eq5}, $\mathbf{R}_\mathbf{X}\approx \mathbf{WW}^H$. Therefore,  $(d)$ holds.
Problem $(\mathcal{P} _{1})$ can be further transformed into $(\mathcal{P} _{1.1})$ as follows,
\begin{subequations}\label{eq27}
 \begin{align}
\left( \mathcal{P} _{1.1} \right)~~  & \mathop{\mathrm{maximize}} \limits_{\,\,\mathbf{W}}   &&\frac{\alpha N_r\mathbf{a}\left( \theta _i \right) \mathbf{WW}^H \mathbf{a}\left( \theta _i \right) ^H}{\alpha N_r\sigma _{\mathrm{t}}^{2}+\sigma _{\mathrm{ap}}^{2}}
\\
&~\mathrm{subject~to}   &&\gamma _{\mathrm{u}}\geqslant \gamma _{{\mathrm{uth}}}
\\
& &&\mathrm{Tr}\left( \mathbf{WW}^H \right) \leqslant P_T.
\end{align}
\end{subequations}
Introducing an auxiliary variables $q$, we can write the epigraph equivalent \cite{boyed2004} form as
\begin{subequations}\label{eq28}
 \begin{align}
\left( \mathcal{P} _{1.2} \right)~~  & \mathop{\mathrm{maximize}} \limits_{\,\,\mathbf{W},~ q}   &&q
\\
&~\mathrm{subject~to}   &&\gamma _{\mathrm{u}}\geqslant \gamma _{{\mathrm{uth}}}
\\
& &&\mathrm{Tr}\left( \mathbf{WW}^H \right) \leqslant P_T
\\
& &&\frac{\alpha N_r\mathbf{a}\left( \theta _i \right) \mathbf{WW}^H \mathbf{a}\left( \theta _i \right) ^H}{\alpha N_r\sigma _{\mathrm{t}}^{2}+\sigma _{\mathrm{ap}}^{2}} \geqslant q
\\
& && q \geqslant 0.
\end{align}
\end{subequations}
Let $\mathbf{R}_\mathbf{W}=\mathbf{WW}^{H}$, 
$\mathbf{W}_\mathrm{u}=\mathbf{w}_\mathrm{u}\mathbf{w}_\mathrm{u}^{H}$, $\mathbf{W}_\mathrm{t}=\mathbf{w}_\mathrm{t}\mathbf{w}_\mathrm{t}^{H}$, $\mathbf{F}=\mathbf{h}_\mathrm{f}^{H}\mathbf{h}_\mathrm{f}$, 
$\mathbf{B}=\mathbf{h}_\mathrm{b}^{H}\mathbf{h}_\mathrm{b}$, 
and $\mathbf{U}=\mathbf{h}_\mathrm{u}^{H}\mathbf{h}_\mathrm{u}$. The UE communication SINR constraint (\ref{eq28}b) can be reformulated as (\ref{eq29}). Note that the channel $\mathbf{h}_\mathrm{f}$ in this case is regarded as $\mathbf{a}(\theta_i)$, and the $h_\mathrm{tu}$ is regarded as its maximum  value$h_{\mathrm{tu}}^{\text{max}}$. Under this setting, the calculated  $\gamma _{\mathrm{u}}$ is actually a lower bound of the real SINR. The optimization could ensure a robust communication performance. The constraint (\ref{eq28}d) can be reformulated as follows,
\begin{figure*}[b]
\hrulefill
\begin{gather}
\begin{aligned}
&\mathrm{Tr}\left( \mathbf{UW}_{\mathrm{u}} \right) -\gamma _{\mathrm{uth}}\mathrm{Tr}\left( \mathbf{U}\left( \mathbf{R}_{\mathbf{W}}-\mathbf{W}_{\mathrm{u}} \right) \right) -\gamma _{\mathrm{uth}}\alpha |h_{\mathrm{tu}}|^2\mathrm{Tr}\left( \mathbf{FR}_{\mathbf{W}} \right) -\gamma _{\mathrm{uth}}\alpha |h_{\mathrm{tu}}|^2\sigma _{\mathrm{t}}^{2}-\gamma _{\mathrm{uth}}\sigma _{\mathrm{u}}^{2}\geqslant 0.
\label{eq29}
\end{aligned}
\end{gather}
\end{figure*}
\begin{gather}
\begin{aligned}
\alpha N_r\mathbf{a}\left( \theta _i \right) \mathbf{R}_{\mathbf{W}}\mathbf{a}\left( \theta _i \right) ^H-q\alpha N_r\sigma _{\mathrm{t}}^{2}-q\sigma _{\mathrm{ap}}^{2}\geqslant 0.
\label{eq30}
\end{aligned}
\end{gather}
Finally, Problem $(\mathcal{P} _{1})$ can be  transformed as 
\begin{subequations}\label{eq31}
 \begin{align}
\left( \mathcal{P} _{1.3} \right)~~  & \mathop{\mathrm{minimize}} \limits_{\,\,\mathbf{R}_{\mathbf{W}},\mathbf{W}_{\mathrm{u}},\mathbf{W}_{\mathrm{t}},q}   &&-q
\\
&~~~\mathrm{subject~to}   &&\eqref{eq29} \nonumber\\
& && \eqref{eq30} \nonumber
\\
& &&\mathrm{Tr}\left( \mathbf{R}_{\mathbf{W}} \right) \leqslant P_T
\\
& && q \geqslant 0
\\
& && \mathbf{R}_{\mathbf{W}}-\mathbf{W}_{\mathrm{u}}-\mathbf{W}_{\mathrm{t}}\succeq 0
\label{eq31d}
\\
& &&\mathrm{rank}\left( \mathbf{W}_{\mathrm{u}} \right) =1
\label{eq31e}
\\
& &&\mathrm{rank}\left( \mathbf{W}_{\mathrm{t}} \right) =1.
\label{eq31f}
\end{align}
\end{subequations}
Due to the rank constraints, optimization problem $(\mathcal{P}_{1.3})$ is still a non-convex problem and difficult to solve. We could use the following SDR  technique\cite{LiuX2020,Luo2010} to obtain the optimal rank-1 solution. In particular, we ignore the two rank constraints (\ref{eq31}e) and (\ref{eq31}f), the problem then becomes a convex semidefinite programming (SDP) problem, which can be solved using the convex optimization tool box, such as CVX. 
We denote the optimal solution of problem $(\mathcal{P} _{1.3})$ ignoring rank constraints as $\overline{\mathbf{R}}_\mathbf{W}$, $\overline{\mathbf{W}}_{\mathrm{u}}$, $\overline{\mathbf{W}}_{\mathrm{t}}$, and $\overline{q}$.  
Next, we could get the rank-1 optimal solution via the following proposition.
\begin{proposition}
Given an optimal solution $\overline{\mathbf{R}}_\mathbf{W}$, $\overline{\mathbf{W}}_{\mathrm{u}}$, $\overline{\mathbf{W}}_{\mathrm{t}}$, and $\overline{q}$ of $(\mathcal{P} _{1.3})$ without rank-1 constrains, the following $\widetilde{\mathbf{R}}_\mathbf{W}$, $\widetilde{\mathbf{W}}_{\mathrm{u}}$, $\widetilde{\mathbf{W}}_{\mathrm{t}}$, and $\widetilde{q}$ is the rank-1 optimal solution of $(\mathcal{P} _{1.3})$
\begin{align}
\label{eq32}
\widetilde{\mathbf{R}}_\mathbf{W}&=\overline{\mathbf{R}}_\mathbf{W}\nonumber\\
\widetilde{\mathbf{W}}_{\mathrm{u}}&=\frac{\overline{\mathbf{W}}_{\mathrm{u}}\mathbf{U}\overline{\mathbf{W}}_{\mathrm{u}}^{H}}{\mathrm{Tr}\left( \mathbf{U}\overline{\mathbf{W}}_{\mathrm{u}} \right)}\nonumber\\
\widetilde{\mathbf{W}}_{\mathrm{t}}&=\frac{\overline{\mathbf{W}}_{\mathrm{t}}\mathbf{F}\overline{\mathbf{W}}_{\mathrm{t}}^{H}}{\mathrm{Tr}\left( \mathbf{F}\overline{\mathbf{W}}_{\mathrm{t}} \right)}\nonumber\\
\widetilde{q}&=\overline{q}.
\end{align}
\label{lm1}
\end{proposition}
\begin{proov}
See Appendix \ref{AP1}.
\end{proov}
\begin{Remark}
It can be seen that even if there are rank constraints in the problem $(\mathcal{P} _{1.3})$, we can first ignore the rank-1 constraints and obtain a solution to the relaxed problem, and then directly obtain a rank-1 global optimal solution according to the Proposition \ref{lm1}. Note that Proposition 1 can be considered as a corollary of Theorem 1 of \cite{LiuX2020} in the B-ISAC system.
\end{Remark}
Next, we calculate the joint beamforming matrix $\mathbf{W}$ via the following formulas 
\begin{align}\label{eq33}
&\widetilde{\mathbf{w}}_{\mathrm{u}}=\left( \mathbf{h}_{\mathrm{u}}\widetilde{\mathbf{W}}_{\mathrm{u}} \mathbf{h}_{\mathrm{u}}^{H} \right) ^{-1/2}\widetilde{\mathbf{W}}_{\mathrm{u}}\mathbf{h}_{\mathrm{u}}^{H}\nonumber
\\
&\widetilde{\mathbf{w}}_{\mathrm{t}}=\left( \mathbf{h}_{\mathrm{f}}\widetilde{\mathbf{W}}_{\mathrm{t}} \mathbf{h}_{\mathrm{f}}^{H} \right) ^{-1/2}\widetilde{\mathbf{W}}_{\mathrm{t}}\mathbf{h}_{\mathrm{f}}^{H}\nonumber
\\
&\widetilde{\mathbf{W}}_{\mathrm{s}}=\mathrm{chol}\left(\widetilde{\mathbf{R}}_{\mathbf{W}}-\widetilde{\mathbf{w}}_{\mathrm{u}}-\widetilde{\mathbf{w}}_{\mathrm{t}}\right)\nonumber
\\
&\mathbf{W}=\left[\widetilde{\mathbf{w}}_{\mathrm{u}}, \widetilde{\mathbf{w}}_{\mathrm{t}}, \widetilde{\mathbf{W}}_{\mathrm{s}} \right].
\end{align}
We summarize the optimal SDR method solving $(\mathcal{P} _{1})$ in Algorithm \ref{alg1}.
\subsection{Joint Beamforming Scheme for Tag Estimation}\label{3B}
In addition to obtain the existence information of the tag (detection), we must also obtain the information about the tag position (``channel''), so we further study the beamforming problem optimizing the estimation performance. In this case, the B-ISAC system should work in tag estimation stage. We considered two
cases of estimation optimization, non-Bayesian (LS) case, and Bayesian (LMMSE) case.
\renewcommand{\algorithmicrequire}{ \textbf{Input:}} 
\renewcommand{\algorithmicensure}{ \textbf{Output:}} 
\begin{algorithm}[!t]
    \caption{Optimal SDR method for solving $(\mathcal{P} _{1})$ $(\mathcal{P} _{2})$ $(\mathcal{P} _{3})$ }
    \label{alg1}
    \begin{algorithmic} [1]
    \REQUIRE \, 
    $P_T$, $\sigma_\mathrm{ap}^2$, $\sigma_\mathrm{t}^2$, $\sigma_\mathrm{u}^2$, $\mathbf{h}_\mathrm{u}$, $N_t$, $N_r$, $\alpha$, $\gamma_\mathrm{uth}$, $\theta_i$($\mathcal{P} _{1}$),  $\theta_{\text{max}}$($\mathcal{P} _{2}$, $\mathcal{P} _{3}$), $\mathbf{R}_\mathbf{G}$($\mathcal{P} _{3}$).       \\
    \ENSURE Designed joint beamforming matrix $\mathbf{W}^{\star}$.
    \renewcommand{\algorithmicensure}{ \textbf{Steps:}}
    \ENSURE \, 
    \STATE Solve \eqref{eq31}/\eqref{eq35}/\eqref{eq37} without rank-1 constrains using convex tool box to obtain $\overline{\mathbf{R}}_\mathbf{W}$, $\overline{\mathbf{W}}_{\mathrm{u}}$, $\overline{\mathbf{W}}_{\mathrm{t}}$.
    \STATE Obtain rank-1 optimal solution $\widetilde{\mathbf{R}}_\mathbf{W}$, $\widetilde{\mathbf{W}}_{\mathrm{u}}$, $\widetilde{\mathbf{W}}_{\mathrm{t}}$ via 
    \eqref{eq32}.
    \STATE Obtain joint beamforming matrix $\mathbf{W}$ via \eqref{eq33}.
    \STATE Return $\mathbf{W}^{\star}=\mathbf{W}$. 
\end{algorithmic}
\end{algorithm}

\textbf{LS Estimation}:
First, assuming that there is no prior information about $\mathbf{G}$, we consider the optimization problem of LS estimation performance. The optimization problem is given by
\begin{subequations}
 \begin{align}
\left( \mathcal{P} _2 \right)~~  & \mathop{\mathrm{minimize}} \limits_{\,\,\mathbf{W}}   &&J_{\mathrm{LS}} 
\\
&~\mathrm{subject~to}   &&\gamma _{\mathrm{u}}\geqslant \gamma _{{\mathrm{uth}}}
\\
& &&\mathrm{Tr}\left( \mathbf{WW}^H \right) \leqslant P_T,
\end{align}
\end{subequations}
where $J_{\mathrm{LS}} $ is the LS estimation error, which is given in \eqref{eq18}. Similar to $\left( \mathcal{P} _1 \right)$, the problem $\left( \mathcal{P} _2 \right)$ can be transformed as
\begin{subequations}\label{eq35}
 \begin{align}
\left( \mathcal{P} _{2.1} \right)~~  & \mathop{\mathrm{minimize}} \limits_{\,\,\mathbf{R}_{\mathbf{W}},\mathbf{W}_{\mathrm{u}},\mathbf{W}_{\mathrm{t}}}   &&J_{\mathrm{LS}}
\\
&~~~\mathrm{subject~to}   &&\eqref{eq29} \nonumber
\\
& &&\mathrm{Tr}\left( \mathbf{R}_{\mathbf{W}} \right) \leqslant P_T
\\
& && \mathbf{R}_{\mathbf{W}}-\mathbf{W}_{\mathrm{u}}-\mathbf{W}_{\mathrm{t}}\succeq 0
\\
& &&\mathrm{rank}\left( \mathbf{W}_{\mathrm{u}} \right) =1
\\
& &&\mathrm{rank}\left( \mathbf{W}_{\mathrm{t}} \right) =1,
\end{align}
\end{subequations}
where $J_{\mathrm{LS}}=\frac{N_r\left( \sigma _{\mathrm{ap}}^{2}+\alpha \left\| \mathbf{h}_{\mathrm{b}} \right\| ^2\sigma _{\mathrm{t}}^{2} \right)}{\alpha L}\mathrm{Tr}\left\{ \left( \mathbf{R}_{\mathbf{W}}\right) ^{-1} \right\} $. 
Note that although we do not know the specific information of $\mathbf{h}_\mathrm{f}$ in this stage, we can assume that after angle scanning through the detection process, we already know the angle direction of the greatest interference $\theta_{\text{max}}$. We set $\mathbf{h}_\mathrm{f}=\mathbf{a}(\theta_{\text{max}})$ and set $h_\mathrm{tu}$ to its maximum value $h_{\mathrm{tu}}^{\text{max}}$ to ensure the robust communication performance. The problem $\left( \mathcal{P} _{2.1} \right)$ can be solved by using the optimal SDR method as Algorithm \ref{alg1}. 

\textbf{LMMSE Estimation}: Second, we consider the optimization of Bayesian estimation performance, when the AP has obtained partial information of $\mathbf{G}$, such as known $\mathbf{R}_{\mathbf{G}}=\mathbb{E} \left\{ \mathbf{G}^H\mathbf{G} \right\}$. In this case, We can consider the asymptotic performance optimization of LMMSE. The optimization problem can be expressed as follows,
\begin{subequations}
 \begin{align}
\left( \mathcal{P} _3 \right)~~  & \mathop{\mathrm{minimize}} \limits_{\,\,\mathbf{W}} 
&&J_{\mathrm{LMMSE}} 
\\
&~\mathrm{subject~to}   &&\gamma _{\mathrm{u}}\geqslant \gamma _{{\mathrm{uth}}}
\\
& &&\mathrm{Tr}\left( \mathbf{WW}^H \right) \leqslant P_T.
\end{align}
\end{subequations}
Similar to $\left( \mathcal{P} _1 \right)$ and $\left( \mathcal{P}_2 \right)$, the problem can be transformed as 
\begin{subequations}\label{eq37}
 \begin{align}
\left( \mathcal{P} _{3.1} \right)~~  & \mathop{\mathrm{minimize}} \limits_{\,\,\mathbf{R}_{\mathbf{W}},\mathbf{W}_{\mathrm{u}},\mathbf{W}_{\mathrm{t}}}   &&J_{\mathrm{LMMSE}}
\\
&~~~\mathrm{subject~to}   &&\eqref{eq29} \nonumber
\\
& &&\mathrm{Tr}\left( \mathbf{R}_{\mathbf{W}} \right) \leqslant P_T
\\
& && \mathbf{R}_{\mathbf{W}}-\mathbf{W}_{\mathrm{u}}-\mathbf{W}_{\mathrm{t}}\succeq 0
\\
& &&\mathrm{rank}\left( \mathbf{W}_{\mathrm{u}} \right) =1
\\
& &&\mathrm{rank}\left( \mathbf{W}_{\mathrm{t}} \right) =1,
\end{align}
\end{subequations}
where $J_{\mathrm{LMMSE}}=\mathrm{Tr}\left\{ \left( \mathbf{R}_{\mathbf{G}}^{-1}+\frac{\alpha L}{N_r\left( \sigma _{\mathrm{ap}}^{2}+\alpha \left\| \mathbf{h}_{\mathrm{b}} \right\| ^2\sigma _{\mathrm{t}}^{2} \right)}{\mathbf{R}_{\mathbf{W}}}\right) ^{-1} \right\}$. The problem $\left( \mathcal{P} _{3.1} \right)$ can be solved by using the optimal SDR method as Algorithm \ref{alg1}. 

{\subsection{Joint Beamforming Scheme for Communication Enhancement}\label{3C}

In communication enhancement stage, we consider the case that the AP has detected the tag and has obtained the complete information about the channel $\mathbf{G}=\mathbf{h}_{\mathrm{b}}\mathbf{h}_{\mathrm{f}}$. We focus on the beamforming design for optimizing the communication rate of legacy communication UE. We consider the problem of maximizing the communication rate of communication UE under the constraints of energy budget, SINRs at the tag and AP. The problem is given by
\begin{subequations}
 \begin{align}
\left( \mathcal{P} _4 \right)~~  & \mathop{\mathrm{maximize}} \limits_{\,\,\mathbf{W}}   &&\log _2\left( 1+\gamma _{\mathrm{u}} \right) 
\\
&~\mathrm{subject~to}   &&\gamma _{\mathrm{t}}\geqslant \gamma _{\mathrm{tth}}
\label{eq38b}
\\
& &&\gamma _{\mathrm{ap}}\geqslant \gamma _{\mathrm{apth}}
\label{eq38c}
\\
& &&\mathrm{Tr}\left( \mathbf{WW}^H \right) \leqslant P_T.
\end{align}
\end{subequations}
This problem is a non-convex optimization problem and requires appropriate transformation to be solved. Firstly, we handle the objective function. Note that since only a single communication UE is considered, optimizing the rate and optimizing the communication SINR $\gamma _{\mathrm{u}}$ are equivalent. However, according to \eqref{eq24}, $\gamma _{\mathrm{u}}$ has a complicated fractional form. By introducing an auxiliary variables $y$, we can convert  the objective function into polynomial form by quadratic transform\cite{Shen2018}. The new objective function $\mathcal{F} \left( \mathbf{W},y \right)$ is given by \eqref{eq39}. Then the optimization problem can be written as
\begin{figure*}[ht]
\begin{gather}
\begin{aligned}
\mathcal{F} \left( \mathbf{W},y \right) =2y\mathrm{Re}\left\{ \mathbf{h}_{\mathrm{u}}\mathbf{w}_{\mathrm{u}} \right\} -y^2\left[ |\mathbf{h}_{\mathrm{u}}\mathbf{w}_{\mathrm{t}}|^2+\sum_{i=1}^{N_t}{|\mathbf{h}_{\mathrm{u}}\mathbf{w}_i|^2}+\alpha |h_{\mathrm{tu}}|^2\left( |\mathbf{h}_{\mathrm{f}}\mathbf{w}_{\mathrm{u}}|^2+|\mathbf{h}_{\mathrm{f}}\mathbf{w}_{\mathrm{t}}|^2+\sum_{i=1}^{N_t}{|\mathbf{h}_{\mathrm{f}}\mathbf{w}_i|^2}+\sigma _{\mathrm{t}}^{2} \right) +\sigma _{\mathrm{u}}^{2} \right] .
\label{eq39}
\end{aligned}
\end{gather}
\vspace{-0.5cm}
\end{figure*}
\begin{subequations}
 \begin{align}
\left( \mathcal{P} _{4.1} \right)~~  & \mathop{\mathrm{maximize}} \limits_{\,\,\mathbf{W},y}   &&\mathcal{F} \left( \mathbf{W},y \right)
\\
&~\mathrm{subject~to}   &&\gamma _{\mathrm{t}}\geqslant \gamma _{\mathrm{tth}}
\label{eq40b}
\\
& &&\gamma _{\mathrm{ap}}\geqslant \gamma _{\mathrm{apth}}
\label{eq40c}
\\
& &&\mathrm{Tr}\left( \mathbf{WW}^H \right) \leqslant P_T,
\end{align}
\end{subequations}
where $\mathcal{F} \left( \mathbf{W},y \right)$  is a conditionally concave
function with respect to each variable given the other. Therefore, we develop an alternating optimization method to solve this problem.

\textbf{Update $y$}: Given $\mathbf{W}$, the optimization for the auxiliary variables $y$ is a convex problem without constraints, given as
\begin{gather}
\begin{aligned}
\label{eq41}
\left( \mathcal{P} _{4.1.1} \right)~~  & \mathop{\mathrm{maximize}} \limits_{\,\,y}   &&\mathcal{F} \left( \mathbf{W},y \right).
\end{aligned}
\end{gather}
Its optimal solution can be obtained straightforwardly by setting $\frac{\partial \mathcal{F}}{\partial y}=0$. The optimal $y^*$ is given by \eqref{eq42}.
\begin{figure*}
\begin{gather}
\begin{aligned}
y^*=\frac{\mathrm{Re}\left\{ \mathbf{h}_{\mathrm{u}}\mathbf{w}_{\mathrm{u}} \right\}}{|\mathbf{h}_{\mathrm{u}}\mathbf{w}_{\mathrm{t}}|^2+\sum_{i=1}^{N_t}{|\mathbf{h}_{\mathrm{u}}\mathbf{w}_i|^2}+\alpha |h_\mathrm{tu}|^2\left( |\mathbf{h}_{\mathrm{f}}\mathbf{w}_{\mathrm{u}}|^2+|\mathbf{h}_{\mathrm{f}}\mathbf{w}_{\mathrm{t}}|^2+\sum_{i=1}^{N_t}{|\mathbf{h}_{\mathrm{f}}\mathbf{w}_i|^2}+\sigma _{\mathrm{t}}^{2} \right) +\sigma _{\mathrm{u}}^{2}}.
\label{eq42}
\end{aligned}
\end{gather}
\hrulefill
\end{figure*}

\textbf{Update $\mathbf{W}$}: Given $y$, the optimization problem can be expressed as 
\begin{subequations}
 \begin{align}
\left( \mathcal{P} _{4.1.2} \right)~~  & \mathop{\mathrm{maximize}} \limits_{\,\,\mathbf{W}=\left[\mathbf{w}_{\mathrm{u}},\mathbf{w}_{\mathrm{t}},\mathbf{w}_{1},\ldots,\mathbf{w}_{N_t} \right] }   &&\mathcal{F} \left( \mathbf{W},y \right)
\\
&~~~~~~~~\mathrm{subject~to}   &&\gamma _{\mathrm{t}}\geqslant \gamma _{\mathrm{tth}}
\label{eq43b}
\\
& &&\gamma _{\mathrm{ap}}\geqslant \gamma _{\mathrm{apth}}
\label{eq43c}
\\
& &&\mathrm{Tr}\left( \mathbf{WW}^H \right) \leqslant P_T.
\end{align}
\end{subequations}
Note that the objective function is concave with respect to $\mathbf{W}$. The main challenge is the non-convex constraints  (\ref{eq43b}) and (\ref{eq43c}). According to \eqref{eq8}, the
(\ref{eq43b}) can be rewritten as
\begin{gather}
\begin{aligned}
\label{eq44}
\frac{1}{\gamma _{\mathrm{tth}}}|\mathbf{h}_{\mathrm{f}}\mathbf{w}_{\mathrm{t}}|^2\geqslant |\mathbf{h}_{\mathrm{f}}\mathbf{w}_{\mathrm{u}}|^2+\sum_{i=1}^{N_t}{|\mathbf{h}_{\mathrm{f}}\mathbf{w}_i|^2}+\sigma _{\mathrm{t}}^{2}.
\end{aligned}
\end{gather}
Take root square of both side of \eqref{eq44}, the constraint becomes a second-order cone form, which is given by
\begin{gather}
\begin{aligned}
\label{eq45}
\sqrt{\frac{1}{\gamma _{\mathrm{tth}}}}\mathrm{Re}\left\{ \mathbf{h}_{\mathrm{f}}\mathbf{w}_{\mathrm{t}} \right\} \geqslant \left\| \begin{array}{c}
	\mathbf{h}_{\mathrm{f}}\mathbf{w}_{\mathrm{u}}\\
	\mathbf{h}_{\mathrm{f}}\mathbf{w}_1\\
	\mathbf{h}_{\mathrm{f}}\mathbf{w}_2\\
	\vdots\\
	\mathbf{h}_{\mathrm{f}}\mathbf{w}_{N_t}\\
	\sigma _{\mathrm{t}}\\
\end{array} \right\| .
\end{aligned}
\end{gather}
The original form of left hand side of \eqref{eq44} after taking root square is $\sqrt{\frac{1}{\gamma _{\mathrm{tth}}}}| \mathbf{h}_{\mathrm{f}}\mathbf{w}_{\mathrm{t}}|$. Note that by multiplying $\mathbf{w}_{\mathrm{t}}$ with a complex number on the unit circle, we can always ensure that $\mathbf{h}_{\mathrm{f}}\mathbf{w}_{\mathrm{t}}$ is a positive real number. Therefore, the constraint \eqref{eq44} can be rewritten as \eqref{eq45}. 
 
According to \eqref{eq12}, the constraint \eqref{eq43c} can be rewritten as 
 \begin{gather}
\begin{aligned}
\label{eq46}
\frac{\alpha}{\gamma _{\mathrm{apth}}}|\mathbf{w}_{\mathrm{r}}\mathbf{h}_{\mathrm{b}}|^2  
\mathrm{Tr}\left( \mathbf{F}\mathbf{WW}^H \right) \geqslant \left\| \begin{array}{c}
	 \sqrt{\alpha}|\mathbf{w}_{\mathrm{r}}\mathbf{h}_{\mathrm{b}}|\sigma _{\mathrm{t}}\\
	\left\| \mathbf{w}_{\mathrm{r}} \right\|\sigma_{\mathrm{ap}}\\
\end{array} \right\|^2 .
\end{aligned}
\end{gather}
\renewcommand{\algorithmicrequire}{ \textbf{Input:}} 
\renewcommand{\algorithmicensure}{ \textbf{Output:}} 
\begin{algorithm}[!t]
    \caption{SCA based algorithm for solving $(\mathcal{P}_{4.1.2.1})$ }
    \label{alg2}
    \begin{algorithmic} [1]
    \REQUIRE \, 
    $P_T$, $\sigma_\mathrm{ap}^2$, $\sigma_\mathrm{t}^2$, $\sigma_\mathrm{u}^2$, $\mathbf{h}_\mathrm{u}$, $\mathbf{h}_\mathrm{f}$, $\mathbf{h}_\mathrm{b}$, $N_t$, $N_r$, $\alpha$, $\gamma_\mathrm{uth}$, $\gamma_\mathrm{tth}$, $\gamma_\mathrm{apth}$, 
    $y$.
    \ENSURE Designed beamforming matrix $\mathbf{W}^{\star}$.
    \renewcommand{\algorithmicensure}{ \textbf{Steps:}}
    \ENSURE \, 
   \STATE Initialize: $\mathbf{W}$, $\delta_\mathrm{th}$, ${I}_{\text{max}}$ , $i=1$, 
    $\delta=\infty$.
   
    \STATE \textbf{while} $i \leqslant {I}_{\text{max}}$ and  $\delta \geqslant \delta_\mathrm{th}$  \textbf{do}
    \STATE ~~~~~Let $\mathbf{W}^{\ddagger}=\mathbf{W}$
    \STATE ~~~~~Update $\mathbf{W}$ by solving  \eqref{eq50}.
    \STATE ~~~~~~ $\delta=|\mathrm{Tr}\left[ \mathbf{W}^H\mathbf{F}(\mathbf{W}-\mathbf{W}^{\ddagger})+\mathbf{W}^T\mathbf{F}^T(\mathbf{W}-\mathbf{W}^{\ddagger})^* \right]|$.
    \STATE ~~~~~$i=i+1$.
    \STATE \textbf{end while}
    \STATE Return $\mathbf{W}^{\star}=\mathbf{W}$. 
\end{algorithmic}
\end{algorithm}

However, the square root of both sides of the constraint \eqref{eq46} cannot be transformed into an affine constraint. Now the problem $\left( \mathcal{P} _{4.1.2} \right)$ can be expressed as
\begin{subequations}
 \begin{align}
\left( \mathcal{P} _{4.1.2.1} \right)~~  & \mathop{\mathrm{maximize}} \limits_{\,\,\mathbf{W}}   &&\mathcal{F} \left( \mathbf{W},y \right)
\\
&~\mathrm{subject~to}   &&\eqref{eq45} \nonumber
\\
& &&\eqref{eq46} \nonumber
\\
& &&\mathrm{Tr}\left( \mathbf{WW}^H \right) \leqslant P_T.
\end{align}
\end{subequations}
The main limitation in solving this problem is that constraint \eqref{eq46} is not an affine constraint. Therefore, we adopt successive convex approximation (SCA) based method \cite{Scutari2014} to solve it. Given a $\mathbf{W}^{\ddagger}$, the convex approximation of \eqref{eq46} at $\mathbf{W}^{\ddagger}$ is given by \eqref{eq49} \cite{Hjørungnes2011}. 
\begin{figure*}
 \begin{gather}
\begin{aligned}
\label{eq49}
\frac{\alpha}{\gamma _{\mathrm{apth}}}|\mathbf{w}_{\mathrm{r}}\mathbf{h}_{\mathrm{b}}|^2 \mathrm{Tr}\left( \mathbf{F}\mathbf{W}^{\ddagger}\mathbf{W}^{\ddagger H} \right)+\frac{\alpha}{\gamma _{\mathrm{apth}}}|\mathbf{w}_{\mathrm{r}}\mathbf{h}_{\mathrm{b}}|^2\mathrm{Tr}\left[ \mathbf{W}^H\mathbf{F}(\mathbf{W}-\mathbf{W}^{\ddagger})+\mathbf{W}^T\mathbf{F}^T(\mathbf{W}-\mathbf{W}^{\ddagger})^* \right]\geqslant &\left\| \begin{array}{c} \sqrt{\alpha}|\mathbf{w}_{\mathrm{r}}\mathbf{h}_{\mathrm{b}}|\sigma _{\mathrm{t}}\\
	\left\| \mathbf{w}_{\mathrm{r}} \right\|\sigma_{\mathrm{ap}}\\
\end{array} \right\|^2 .
\end{aligned}
\end{gather}
\hrulefill
\end{figure*}
The convex approximation problem of $\left( \mathcal{P} _{4.1.2.1} \right)$ can be written as follows,
\begin{subequations}
\begin{align}
\left( \mathcal{P} _{4.1.2.2} \right)~~  & \mathop{\mathrm{maximize}} \limits_{\,\,\mathbf{W}}   &&\mathcal{F} \left( \mathbf{W},y \right)
\\
&~\mathrm{subject~to}   &&\eqref{eq45} \nonumber
\\
& &&\eqref{eq49} \nonumber
\\
& &&\mathrm{Tr}\left( \mathbf{WW}^H \right) \leqslant P_T.
\end{align}
\label{eq50}
\end{subequations}
The SCA based method for solving $\left( \mathcal{P}_{4.1.2.1} \right)$ has been briefly summarized in 
Algorithm \ref{alg2}. Based on the above derivations, we can update $y$ and $\mathbf{W}$ iteratively to obtain the joint beamformer. The alternating joint beamforming design for communication rate optimization is summarized in Algorithm \ref{alg3}. With appropriate initialization, we iteratively update each variable until convergence.
\renewcommand{\algorithmicrequire}{ \textbf{Input:}} 
\renewcommand{\algorithmicensure}{ \textbf{Output:}} 
\begin{algorithm}[!t]
    \caption{Alternating algorithm for sloving $(\mathcal{P}_{4})$ }
    \label{alg3}
  \begin{algorithmic} [1]
    \REQUIRE \, 
    $P_T$, $\sigma_\mathrm{ap}^2$, $\sigma_\mathrm{t}^2$, $\sigma_\mathrm{u}^2$, $\mathbf{h}_\mathrm{u}$, $\mathbf{h}_\mathrm{f}$, $\mathbf{h}_\mathrm{b}$, $N_t$, $N_r$, $\alpha$, $\gamma_\mathrm{uth}$, $\gamma_\mathrm{tth}$, $\gamma_\mathrm{apth}$.
    \ENSURE Designed beamforming matrix $\mathbf{W}^{*}$.
    \renewcommand{\algorithmicensure}{ \textbf{Steps:}}
    \ENSURE \, 
   \STATE Initialize: $\mathbf{W}$, $\varepsilon_\mathrm{th}$, ${K}_{\text{max}}$ , $k=1$, 
    $\varepsilon=\infty$, $y_0=0$. 
    
    \STATE \textbf{while} $k \leqslant {K}_{\text{max}}$ and  $\varepsilon \geqslant \varepsilon_\mathrm{th}$  \textbf{do}
    \STATE ~~~~~Update $y_k$ by \eqref{eq42}. 
    \STATE ~~~~~Update $\mathbf{W}$ by using \textbf{Algorithm \ref{alg2}}.
    \STATE ~~~~ $\varepsilon=y_k-y_{k-1}$.
    \STATE ~~~~~$k=k+1$.
    \STATE \textbf{end while}
    \STATE Return $\mathbf{W}^{{*}}=\mathbf{W}$.
\end{algorithmic}
\end{algorithm}

\section{Simulation Results} \label{sec4}
In this section, we present numerical results of the proposed beamforming schemes to verify their effectiveness. We set the transmit and receive array with the same elements number, i.e., $N_t =N_r=16$. We set signal length $L=2048$. We assume that the noise power at the AP, tag, and UE are equal as $\sigma _\mathrm{ap}^{2}=\sigma _\mathrm{t}^{2}=\sigma _\mathrm{u}^{2}=-40$ dBm \cite{Xu2024}. The settings we adopted are close to the parameter settings of actual networks.
\begin{figure}[!t]
\captionsetup{font=small}
\begin{center}
\includegraphics[width=0.45\textwidth, trim=10 1 30 15, clip]{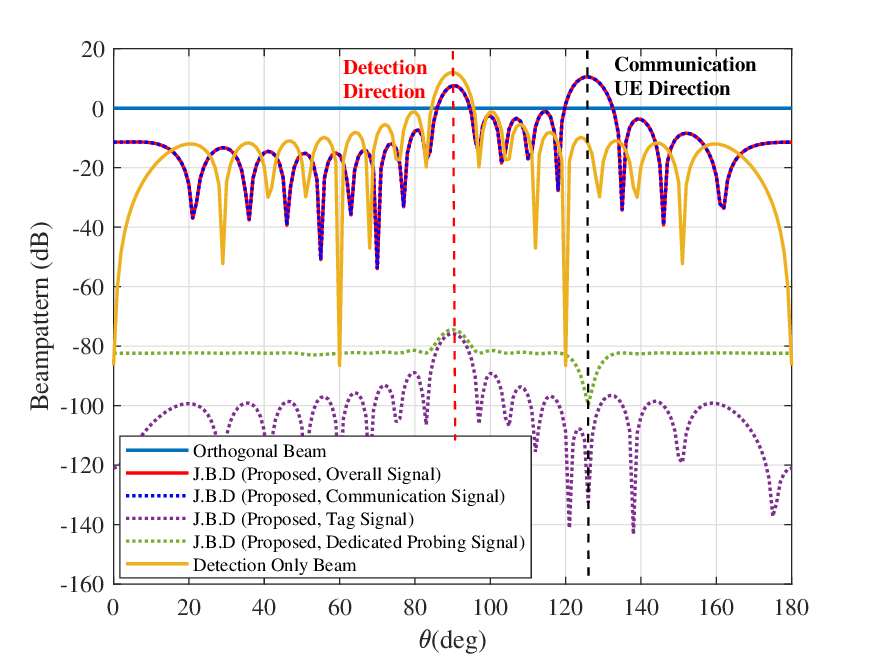}
\end{center}
\caption{Beampattern of joint beamforming scheme for Tag Detection (J.B.D).}
\label{Fig2}
\end{figure}
\begin{figure}[!t]
\centering
\subfigure[$P_D$ of the proposed scheme J.B.D versus UE SINR threshold.]{
\includegraphics[width=0.2276\textwidth]{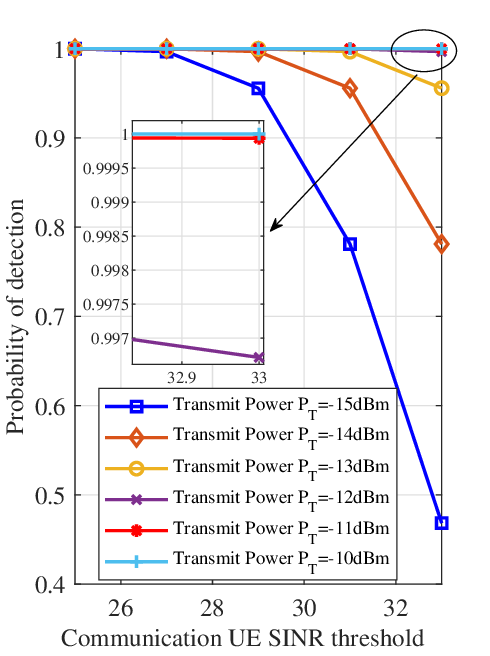}
}
\subfigure[$P_D$ of the proposed scheme J.B.D versus transmit power.]{
\includegraphics[width=0.2276\textwidth]{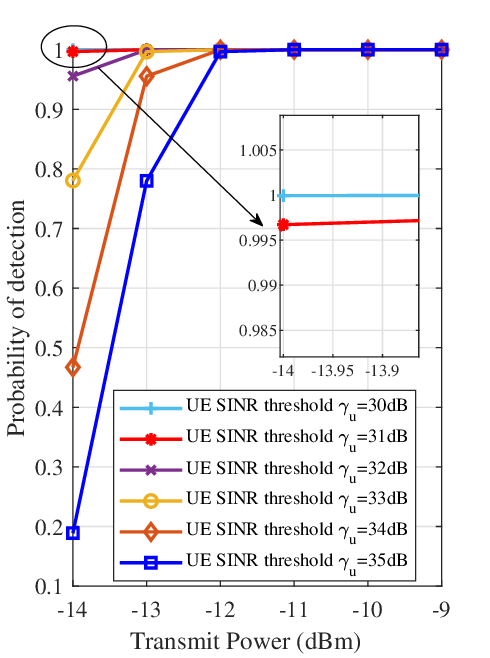}
}
\captionsetup{font=small}
\caption{$P_D$ of the proposed J.B.D scheme versus the UE SINR threshold
and the transmit power.}
\label{Fig3}
\end{figure}

\subsection{Results of Joint Beamforming for Tag Detection}\label{4a}
First, we evaluate the joint beamforming scheme for tag detection. We set the transmit power of the AP as $P_T=0$ dBm, set the maximum value of $h_{\mathrm{tu}}$  as $h_{\mathrm{tu}}^{\text{max}}=0.5$, and set the angle of UE as $\theta_u=126^{\circ}$.  We use the line-of-sight (LOS) channel model in the simulation, which means the corresponding channel $\mathbf{h}_\mathrm{u}$ is $\alpha_\mathrm{u}\mathbf{a}(\theta_u)$. $\alpha_\mathrm{u}$ is channel fading coefficient, which is set as $\alpha_\mathrm{u}=0.8$. The communication UE SINR threshold is set as $\gamma_\mathrm{uth}=15$dB. Fig.~\ref{Fig2} shows the beampattern of the joint beamforming sheme for detection mode (J.B.D) when detecting the $90^{\circ}$ angle grid. As can be seen from Fig.~\ref{Fig2}, in order to ensure both detection and communication performance, the beam pattern forms two high-gain beams in the detection direction and communication UE direction. J.B.D (Proposed, Overall Signal), J.B.D (Proposed, Communication Signal), J.B.D (Proposed, Tag Signal), and J.B.D (Proposed, Dedicated Probing Signal) are beampatterns of the overall signal $\mathbf{X}$, the communication signal $\mathbf{w}_{\mathrm{u}}\mathbf{s}_{\mathrm{u}}^{H}$, the tag signal $\mathbf{w}_{\mathrm{t}}\mathbf{s}_{\mathrm{t}}^{H}$, and the dedicated probing signal $\mathbf{W}_{\mathrm{s}}\mathbf{S}_{\mathrm{s}}$, respectively. Recall that $\mathbf{X}$, $\mathbf{w}_{\mathrm{u}}\mathbf{s}_{\mathrm{u}}^{H}$, $\mathbf{w}_{\mathrm{t}}\mathbf{s}_{\mathrm{t}}^{H}$, and $\mathbf{W}_{\mathrm{s}}\mathbf{S}_{\mathrm{s}}$ represent the overall signal, the communication signal, the tag signal, and the dedicated probing signal, respectively, as defined in \eqref{eq1}. These beampatterns are produced by 
the proposed joint beamforming scheme for tag detection (solving problem  $\left( \mathcal{P} _1 \right)$). Orthogonal Beam is the full orthogonal beamforming scheme with $\mathbf{R}_\mathbf{X}=\frac{P_T}{N_t}\mathbf{I}_{N_t}$, which radiates the same energy at different angles. Detection Only Beam is produced by solving problem $\left( \mathcal{P} _1 \right)$ without the UE communication constraint $\gamma _\mathrm{u}\geqslant \gamma_{\mathrm {uth}}$, which means the scheme only considers the detection performance. This problem can be transformed into a convex problem in the same way as $ \left( \mathcal{P} _1 \right)$, and can be solved easily. In Fig.~\ref{Fig2}, it is worth noting that in order to reduce the additional interference of the detection task on UE communication, the transmit power in this case is mainly allocated to the communication signal, and the energy of the tag signal and the dedicated probing signal are kept as low as possible. Since the transmit signal is known to the AP, the communication signal can also achieve tag detection function, and therefore this design result is reasonable. We have also examined how the $P_D$ varies with the communication SINR threshold and the transmit power.
From Fig.~\ref{Fig3}, we can see that  $P_D$ decreases as the communication performance increases, and it increases as the transmit power increases. This reflects the performance trade-off between UE communication performance and tag detection performance. 
\begin{figure}[!t]
\captionsetup{font=small}
\begin{center}
\includegraphics[width=0.45\textwidth, trim=10 1 30 15, clip]{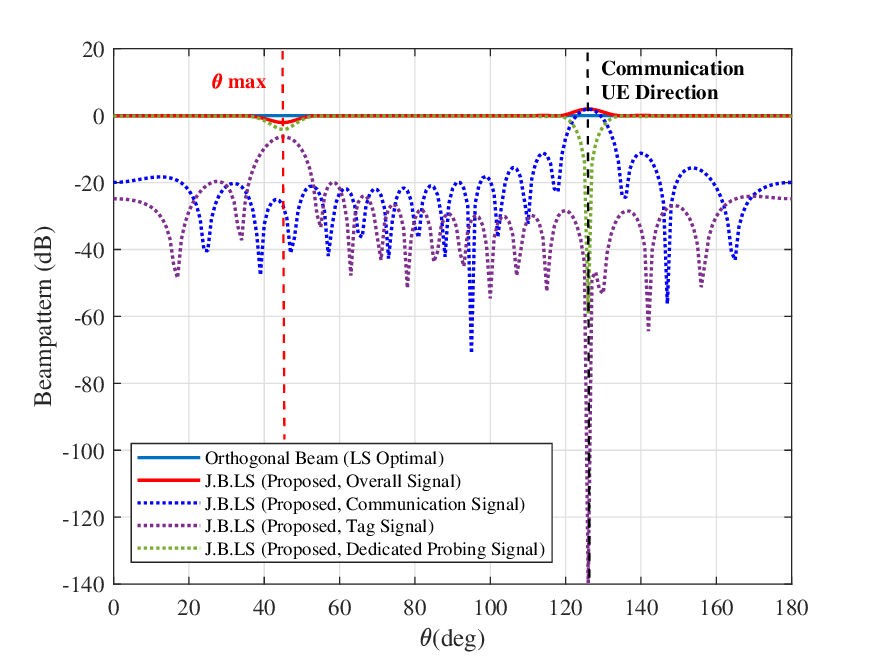}
\end{center}
\caption{Beampattern of joint beamforming scheme for tag estimation by LS (J.B.LS).}
\label{Fig4}
\end{figure}
\begin{figure}[!t]
\captionsetup{font=small}
\begin{center}
\includegraphics[width=0.45\textwidth, trim=10 1 30 15, clip]{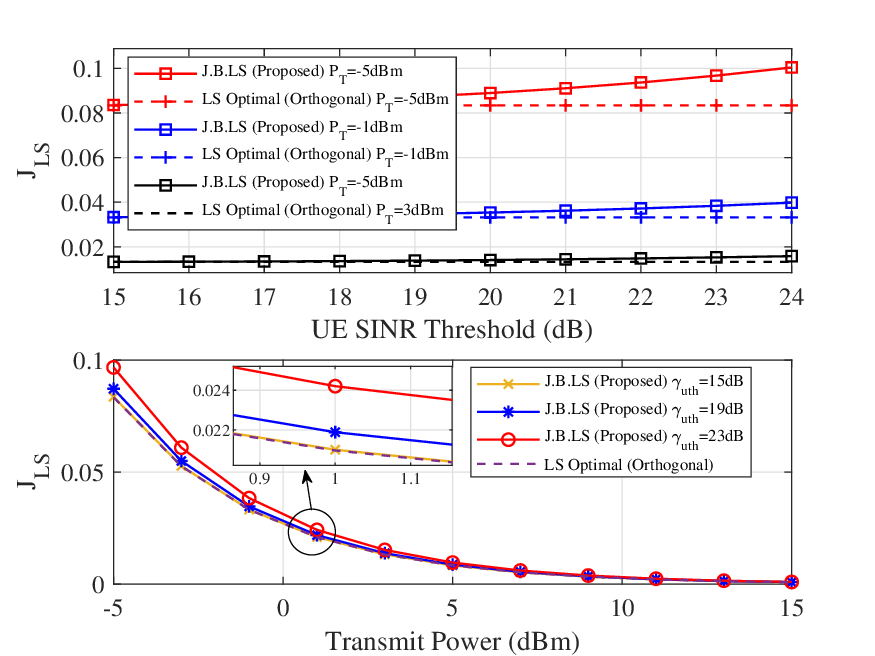}
\end{center}
\caption{LS estimation error of the proposed J.B.LS scheme versus the UE SINR threshold and the transmit power}
\label{Fig5}
\end{figure}
\subsection{Results of Joint Beamforming for Tag Estimation}\label{4b}
Next, we evaluate the joint beamforming scheme for LS estimation (J.B.LS). We set the transmit power of the AP as $P_T=0$ dBm, and set the maximum value of $h_{\mathrm{tu}}^{\text{max}}=0.5$, and set the angle of UE as $\theta_u=126^{\circ}$. We assume that the direction $\theta_\mathrm{max}=45^{\circ}$ with the greatest interference has been obtained after the detection stage. The communication UE SINR threshold is set as $\gamma_\mathrm{uth}=18$dB. We present the beampattern of the proposed J.B.LS in Fig.~\ref{Fig4}. J.B.LS (Proposed, Overall Signal), J.B.LS (Proposed, Communication Signal), J.B.LS (Proposed, Tag Signal), and J.B.LS (Proposed, Dedicated Probing Signal) are beampatterns of the overall
signal, the communication signal, the tag signal, and the dedicated probing signal produced by the proposed J.B.LS scheme (solving problem $\left( \mathcal{P} _2 \right)$) respectively. Orthogonal Beam is the full orthogonal beamforming scheme  with $\mathbf{R}_\mathbf{X}=\frac{P_T}{N_t}\mathbf{I}_{N_t}$. Note that Orthogonal beam is also the optimal for LS estimation according to \eqref{eq19}.
 
In Fig.~\ref{Fig4}, we can see that the proposed J.B.LS scheme forms a huge notch in the communication direction for the tag signal and dedicated probing signal in order to ensure the SINR of UE communication. In this case, the dedicated probing signal will try its best to make the overall transmit signal  $\mathbf{X}$ close to the fully orthogonal optimal signal to optimize the performance of LS estimation.

In Fig.~\ref{Fig5}, we evaluate the estimation error of the proposed J.B.LS scheme versus the UE SINR
threshold and the transmit power. It can be found from Fig.~\ref{Fig5} that as the SINR requirement of the UE communication continues to increase, the estimation error of the proposed scheme J.B.LS will gradually increase, far away from the optimal estimation performance, which reveals the performance trade-off between estimation task and UE communication task. There is no doubt that fully orthogonal beams have the best LS estimation performance. We can also find that as the transmit power increases, the estimation performance of the proposed  J.B.LS scheme will gradually approach the optimal performance, since more transmit power is exploited to improve the estimation performance for a given communication SINR requirement.

We further evaluate the joint beamforming scheme for LMMSE estimation (J.B.LMMSE) with channel prior information. Simulation parameters remain consistent with LS estimation. Given $\mathbf{R}_{\mathbf{G}}$, the beampattern of the proposed J.B.LMMSE scheme is presented in Fig.~\ref{Fig6}. J.B.LMMSE (Proposed, Overall Signal), J.B.LMMSE (Proposed, Communication Signal), J.B.LMMSE (Proposed, Tag Signal), and J.B.LMMSE (Proposed, Dedicated Probing Signal) are beampatterns of the overall signal, the communication signal, the tag signal, and the dedicated probing signal produced by the proposed J.B.LMMSE scheme (solving problem  $\left( \mathcal{P} _3 \right)$) respectively. Orthogonal Beam is the full orthogonal beamforming scheme. The LMMSE optimal scheme is obtained from \eqref{eq22}, which is known as water filling. Similar to the LS situation, J.B.LMMSE scheme forms a notch in the communication UE derection to ensure the SINR of the communication. 

Fig.~\ref{Fig7} shows a partial enlargement of the beampattern. As can be seen from the figure, the beampattern of the proposed scheme not only forms a high-gain beam in the communication direction, but also controls the energy in the $\theta_\mathrm{max}$ direction to avoid excessive interference to communication. On this basis, the J.B.LMMSE beam is as close to the optimal as possible to ensure LMMSE estimation performance.

Fig.~\ref{Fig8} shows the LMMSE estimation error of the proposed J.B.LMMSE scheme versus the
UE SINR threshold and the transmit power. From Fig.~\ref{Fig8}, we can observe the performance trade-off between communication and estimation: the LMMSE error increases as the communication SINR threshold $\gamma_\mathrm{uth}$ increases. At the same time, it can be found that with the fixed communication SINR constraint, as the transmit power increases, the LMMSE estimation performance of the proposed scheme gradually approaches the optimal performance.
\begin{figure}[!t]
\captionsetup{font=small}
\begin{center}
\includegraphics[width=0.46\textwidth, trim=10 1 30 15, clip]{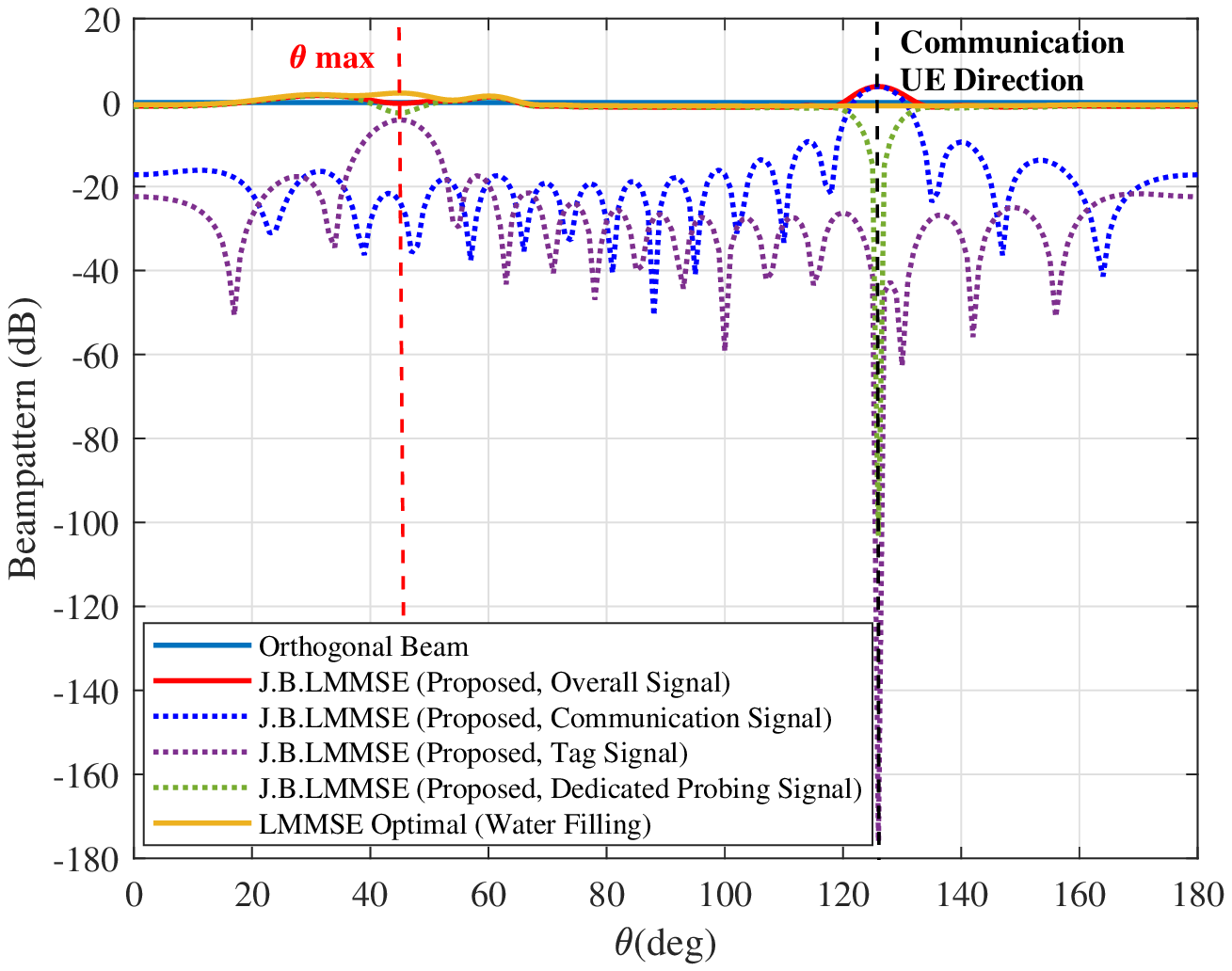}
\end{center}
\caption{Beampattern of joint beamforming scheme for tag estimatio by LMMSE (J.B.LMMSE).}
\label{Fig6}
\end{figure}
\begin{figure}[!t]
\captionsetup{font=small}
\begin{center}
\includegraphics[width=0.46\textwidth, trim=10 1 30 15, clip]{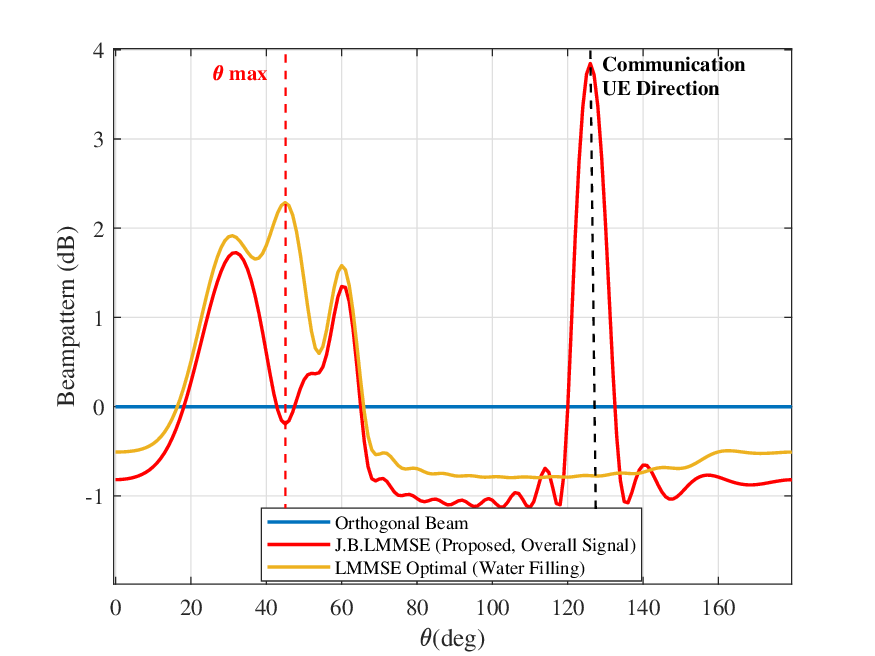}
\end{center}
\caption{Partial enlargement of the beampattern of joint beamforming scheme for tag estimation by LMMSE (J.B.LMMSE).}
\label{Fig7}
\end{figure}
\begin{figure}[!t]
\captionsetup{font=small}
\begin{center}
\includegraphics[width=0.46\textwidth, trim=10 1 30 15, clip]{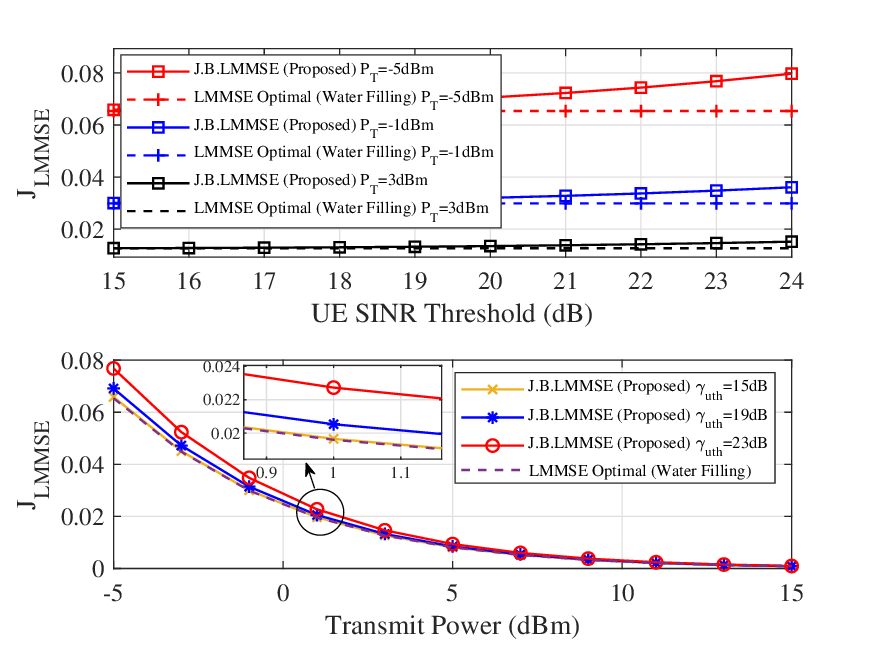}
\end{center}
\caption{LMMSE estimation error of the proposed J.B.LMMSE scheme versus the UE SINR threshold and the transmit power}
\label{Fig8}
\end{figure}
\subsection{Results of Joint Beamforming for  Communication 
 Enhancement}\label{4c}
We analyze the convergence of the proposed Algorithm \ref{alg2} and Algorithm \ref{alg3}. We present the convergence performance of these two algorithms in Fig.~\ref{fig9}(a) and Fig.~\ref{fig9}(b), respectively. The achievable communication rate versus the number of iterations under different setting is present in the figure. Algorithm \ref{alg2} has minor change of rate after 5 iterations and converges after 10 iterations. Algorithm \ref{alg3} has minor change of rate after 2 iterations and converges after 5 iterations. The above simulation results fully demonstrate the quick convergence performance of the proposed algorithms.

Next, we evaluate the beampattern of the proposed joint beamforming scheme for communication rate optimization (J.B.C). We set the transmit power of the AP as $P_T=0$ dBm, and set the $h_{\mathrm{tu}}=0.5$. We use the LOS channel model and set the channels to $\mathbf{h}_\mathrm{f}=0.8\mathbf{a}(\frac{\pi}{4})$, $\mathbf{h}_\mathrm{b}=0.8\mathbf{b}(\frac{\pi}{4})$, and  $\mathbf{h}_\mathrm{u}=0.8\mathbf{a}(\frac{7\pi}{10})$. The SINR threshold at the tag is set to $\gamma_\mathrm{tth}=15$dB to ensure the tag is activated. The SINR threshold at the AP is set to $\gamma_\mathrm{tth}=12$dB to make sure the tag can be detected and communicate with the AP. The beampattern of the proposed J.B.C scheme is presented in Fig.~\ref{Fig10}. J.B.C (Proposed, Overall Signal), J.B.C (Proposed, Communication Signal), J.B.C (Proposed, Tag Signal), and J.B.C (Proposed, Dedicated Probing Signal) are beampatterns of he
overall signal, the communication signal, the tag signal, and the dedicated probing signal produced by the proposed J.B.C scheme (solving problem  $\left( \mathcal{P} _4 \right)$) respectively. Orthogonal Beam is the full orthogonal beamforming scheme. We can observe that the communication beam forms a high gain beam at the communication UE direction and a notch at the tag direction. The tag signal and dedicated probing signal form a high-gain beam in the tag direction and a notch in the communication direction.

The achievable communication rate versus the
transmit power under different settings is depicted in Fig.~\ref{Fig11}. We can also find that as the transmit power increases, the communication performance of the proposed J.B.C scheme increases. We can also observe that the higher the SINR constraints, the lower the achievable communication rate, which reflects the power competition between the tag and UE.
\begin{figure}[t]
\centering
\subfigure[Convergence of Algorithm \ref{alg2}.]{
\includegraphics[width=0.2276\textwidth]{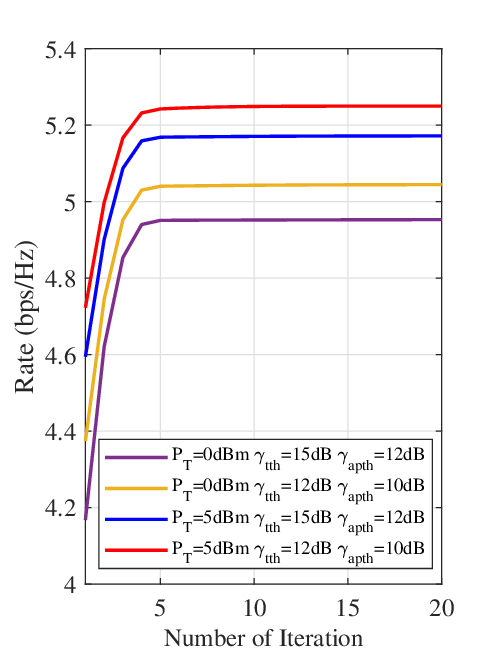}
}
\subfigure[Convergence of Algorithm \ref{alg3}.]{
\includegraphics[width=0.2276\textwidth]{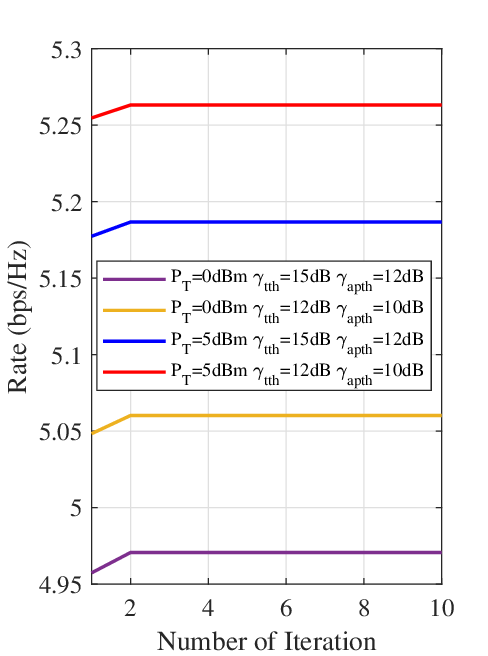}
}
\captionsetup{font=small}
\caption{Convergence performance of the proposed Algorithms.}
\label{fig9}
\end{figure}
\begin{figure}[!t]
\captionsetup{font=small}
\begin{center}
\includegraphics[width=0.46\textwidth, trim=10 1 30 15, clip]{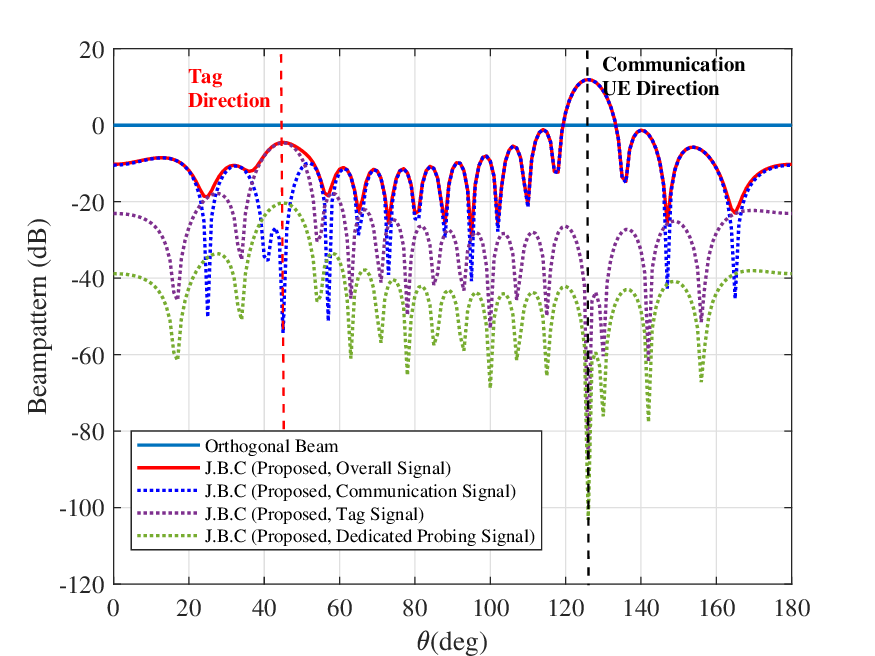}
\end{center}
\caption{Beampattern of joint beamforming scheme for communication enhancement (J.B.C).}
\label{Fig10}
\vspace{-0.3cm}
\end{figure}

\begin{figure}[!t]
\captionsetup{font=small}
\begin{center}
\includegraphics[width=0.46\textwidth, trim=10 1 30 15, clip]{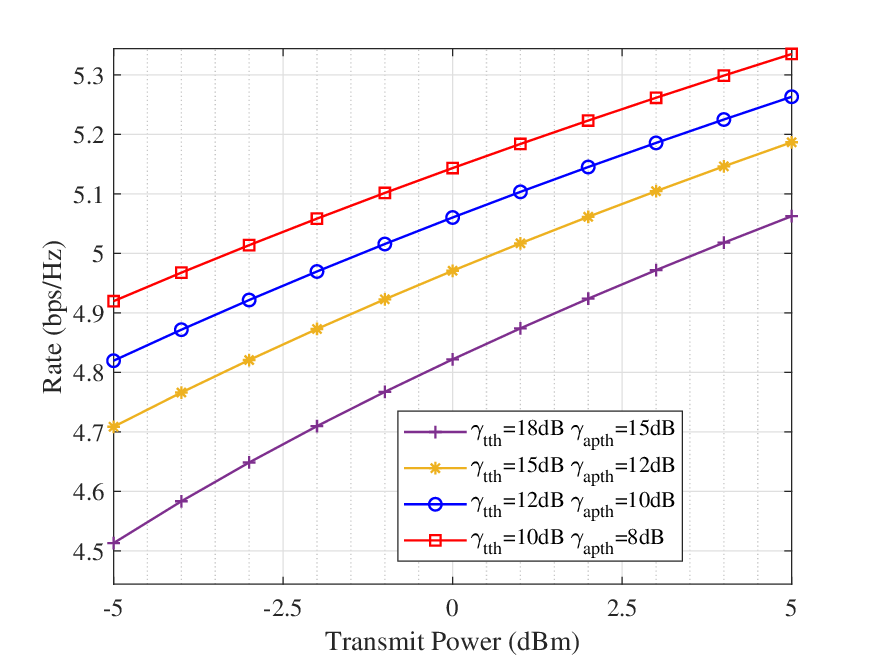}
\end{center}
\caption{The achievable communication rate of the proposed J.B.C scheme versus the
transmit power}
\label{Fig11}
\vspace{-0.3cm} 
\end{figure}
\subsection{Discussion}\label{4d}
Our framework contains three stages (task modes) and therefore is closer to the practical B-ISAC. We have proposed corresponding joint beamforming schemes for these task modes and achieved quite different beampatterns. These will guide the practical system design, and task decomposition and solving. From Figs. \ref{Fig3}, \ref{Fig5}, \ref{Fig8}, and \ref{Fig11}, there is a clear performance trade-off between communication and sensing functions. Therefore, in actual system design, the number of antennas and transmit power must be appropriately designed based on the communication and sensing performance requirements, noise power, and channel conditions.

In our work, we only consider the case of one communication UE and one RF tag, which is a preliminary exploration of the proposed B-ISAC system. When the numbers of communication UEs and RF tags increase, the sensing and communication metrics formulation will change accordingly, due to the presence of multiuser interference. Multiple access technologies can be used to address this issue, however will introduce some new challenges. This interesting topic is beyond the scope of this work and deserves our future study.
}

\section{Conclusion} \label{sec5}
In this paper, we proposed an integrated BackCom and ISAC system called B-ISAC system, in which the passive tag plays a role as a sensing target. The B-ISAC system can enhance the  sensing capabilities while enabling low-power data transmission. We carefully analyzed the communication and sensing performance of the system and designed an optimization framework for joint beamforming of different stages (task modes) in B-ISAC systems. There are three stages in the proposed framework, including tag detection, tag estimation, and communication enhancement. We formulate corresponding optimization problem aiming to enhancing the performance objectives of the respective stages. An SDR optimal algorithm (Algorithm 1) is developed for solving the non-convex optimization problems of tag detection and estimation stages. An iterative algorithm (Algorithm 3) and an SCA algorithm (Algorithm 2) are developed to solve the complicated non-convex problem of communication enhancement stage. Simulation results validate the effectiveness of the proposed algorithms as well as illustrating the performance trade-off between communication and sensing performance. The proposed B-ISAC system has the potential for wide-scale applications in IoE scenarios.
\appendices
\section{Proof of Proposition 1}
\label{AP1}
Given an optimal solution $\overline{\mathbf{R}}_\mathbf{W}$, $\overline{\mathbf{W}}_{\mathrm{u}}$, $\overline{\mathbf{W}}_{\mathrm{t}}$, and $\overline{q}$ of $(\mathcal{P} _{1.3})$ without rank-1 constrains, the $\widetilde{\mathbf{R}}_\mathbf{W}$, $\widetilde{\mathbf{W}}_{\mathrm{u}}$, $\widetilde{\mathbf{W}}_{\mathrm{t}}$, and $\widetilde{q}$ can still achieve the same optimal objective value of $(\mathcal{P} _{1.3})$. This is because the $\widetilde{\mathbf{R}}_\mathbf{W}=\overline{\mathbf{R}}_\mathbf{W}$, the value of the objective function is only related to the variable $\overline{\mathbf{R}}_\mathbf{W}$. Next, we only need to prove that  $\overline{\mathbf{R}}_\mathbf{W}$, $\widetilde{\mathbf{W}}_{\mathrm{u}}$, $\widetilde{\mathbf{W}}_{\mathrm{t}}$, and $\widetilde{q}$ satisfy the constraints of the $(\mathcal{P} _{1.3})$, it can be shown that they are the rank-1 optimal solution.

First, we check the rank-1 constraint. Since $\mathbf{F}$ and $\mathbf{U}$ are rank-1 matrix, the $\widetilde{\mathbf{W}}_{\mathrm{u}}=\frac{\overline{\mathbf{W}}_{\mathrm{u}}\mathbf{U}\overline{\mathbf{W}}_{\mathrm{u}}^{H}}{\mathrm{Tr}\left( \mathbf{U}\overline{\mathbf{W}}_{\mathrm{u}} \right)}$ and $
\widetilde{\mathbf{W}}_{\mathrm{t}}=\frac{\overline{\mathbf{W}}_{\mathrm{t}}\mathbf{F}\overline{\mathbf{W}}_{\mathrm{t}}^{H}}{\mathrm{Tr}\left( \mathbf{F}\overline{\mathbf{W}}_{\mathrm{t}} \right)}$ are easily to be checked as rank-1.

Secondly, we check the power constraint. Since $\widetilde{\mathbf{R}}_\mathbf{W}=\overline{\mathbf{R}}_\mathbf{W}$, $\mathrm{Tr}\left(\widetilde{\mathbf{R}}_\mathbf{W}\right) = \overline{\mathbf{R}}_\mathbf{W} \leqslant P_T$. It means the power constraint is satisfied.

Thirdly, we check the constraint \eqref{eq30}. Since $\widetilde{\mathbf{R}}_\mathbf{W}=\overline{\mathbf{R}}_\mathbf{W}$, $\widetilde{q}=\overline{q}$, then
\begin{align}
&\alpha N_r\mathbf{a}\left( \theta _i \right) \widetilde{\mathbf{R}}_\mathbf{W}\mathbf{a}\left( \theta _i \right) ^H-\widetilde{q}\alpha N_r\sigma _{\mathrm{t}}^{2}-\widetilde{q}\sigma _{\mathrm{ap}}^{2}\nonumber\\
=&\alpha N_r\mathbf{a}\left( \theta _i \right) \overline{\mathbf{R}}_\mathbf{W}\mathbf{a}\left( \theta _i \right) ^H-\overline{q}\alpha N_r\sigma _{\mathrm{t}}^{2}-\overline{q}\sigma _{\mathrm{ap}}^{2}\nonumber\\
\geqslant& 0.
\label{eq51}
\end{align}
The constraint \eqref{eq30} is satisfied.

Then, we check the constraint \eqref{eq29}. Since $\widetilde{\mathbf{W}}_{\mathrm{u}}=\frac{\overline{\mathbf{W}}_{\mathrm{u}}\mathbf{U}\overline{\mathbf{W}}_{\mathrm{u}}^{H}}{\mathrm{Tr}\left( \mathbf{U}\overline{\mathbf{W}}_{\mathrm{u}} \right)}$, we have
\begin{align}
&~~~~ &&\mathrm{Tr}\left( \mathbf{U}\widetilde{\mathbf{W}}_{\mathrm{u}} \right)
=\mathrm{Tr}\left( \mathbf{U}\frac{\overline{\mathbf{W}}_{\mathrm{u}}\mathbf{U}\overline{\mathbf{W}}_{\mathrm{u}}^{H}}{\mathrm{Tr}\left( \mathbf{U}\overline{\mathbf{W}}_{\mathrm{u}} \right)}\right)\overset{\left( a \right)}{=}\mathrm{Tr}\left( \mathbf{U}\overline{\mathbf{W}}_{\mathrm{u}} \right)\nonumber \\
& &&\geqslant\gamma _{\mathrm{uth}}\mathrm{Tr}\left( \mathbf{U}\left( \overline{\mathbf{R}}_{\mathbf{W}} -\overline{\mathbf{W}}_{\mathrm{u}} \right) \right) -\gamma _{\mathrm{uth}}\alpha |h_{\mathrm{tu}}|^2\mathrm{Tr}\left( \mathbf{F}\overline{\mathbf{R}}_{\mathbf{W}}\right)\nonumber\\ 
& &&~~~~-\gamma _{\mathrm{uth}}\alpha |h_{\mathrm{tu}}|^2\sigma _{\mathrm{t}}^{2}-\gamma _{\mathrm{uth}}\sigma _{\mathrm{u}}^{2}\nonumber\\
& &&=\gamma _{\mathrm{uth}}\mathrm{Tr}\left( \mathbf{U}\left( \widetilde{\mathbf{R}}_{\mathbf{W}} -\widetilde{\mathbf{W}}_{\mathrm{u}} \right) \right) -\gamma _{\mathrm{uth}}\alpha |h_{\mathrm{tu}}|^2\mathrm{Tr}\left( \mathbf{F}\widetilde{\mathbf{R}}_{\mathbf{W}}\right)\nonumber\\ 
& &&~~~~-\gamma _{\mathrm{uth}}\alpha |h_{\mathrm{tu}}|^2\sigma _{\mathrm{t}}^{2}-\gamma _{\mathrm{uth}}\sigma _{\mathrm{u}}^{2}\nonumber\\
& &&\geqslant 0.
\label{eq52}
\end{align}
Equation ($a$) holds because the $\mathbf{U}\overline{\mathbf{W}}_{\mathrm{u}}$ is rank-1 matrix. So, the constraint \eqref{eq29} is satisfied

Last, we check the constraint \eqref{eq31d}. We show that $\overline{\mathbf{W}}_{\mathrm{u}}-\widetilde{\mathbf{W}}_{\mathrm{u}}$ and $\overline{\mathbf{W}}_{\mathrm{t}}-\widetilde{\mathbf{W}}_{\mathrm{t}}$ are positive semidefinite matrix. For any $\mathbf{v} \in \mathbb{C} ^{N_t \times 1}$, it holds that
\begin{gather}
\begin{aligned}
\mathbf{v}^{H}(\overline{\mathbf{W}}_{\mathrm{u}}-\widetilde{\mathbf{W}}_{\mathrm{u}})\mathbf{v}=\mathbf{v}^{H}\overline{\mathbf{W}}_{\mathrm{u}}\mathbf{v}-\mathbf{v}^{H}\frac{\overline{\mathbf{W}}_{\mathrm{u}}\mathbf{U}\overline{\mathbf{W}}_{\mathrm{u}}^{H}}{\mathrm{Tr}\left( \mathbf{U}\overline{\mathbf{W}}_{\mathrm{u}} \right)}\mathbf{v}
\label{eq53}
\end{aligned}
\end{gather}
\begin{gather}
\begin{aligned}
\mathbf{v}^{H}(\overline{\mathbf{W}}_{\mathrm{t}}-\widetilde{\mathbf{W}}_{\mathrm{t}})\mathbf{v}=\mathbf{v}^{H}\overline{\mathbf{W}}_{\mathrm{t}}\mathbf{v}-\mathbf{v}^{H}\frac{\overline{\mathbf{W}}_{\mathrm{t}}\mathbf{F}\overline{\mathbf{W}}_{\mathrm{t}}^{H}}{\mathrm{Tr}\left( \mathbf{F}\overline{\mathbf{W}}_{\mathrm{t}} \right)}\mathbf{v}
\end{aligned}
\end{gather}
According to the Cauchy-Schwarz inequality, we have
\begin{gather}
\begin{aligned}
\mathbf{v}^{H}\frac{\overline{\mathbf{W}}_{\mathrm{u}}\mathbf{U}\overline{\mathbf{W}}_{\mathrm{u}}^{H}}{\mathrm{Tr}\left( \mathbf{U}\overline{\mathbf{W}}_{\mathrm{u}} \right)}\mathbf{v} = \frac{\mathbf{v}^{H}\overline{\mathbf{W}}\mathbf{h}_{\mathrm{u}}\mathbf{h}_\mathrm{u}^{H}\overline{\mathbf{W}}_{\mathrm{u}}^{H}\mathbf{v} }{\mathrm{Tr}\left( \mathbf{U}\overline{\mathbf{W}}_{\mathrm{u}}\right)}
\leqslant \frac{\mathbf{v}^{H}\overline{\mathbf{W}}_{\mathrm{u}}\mathbf{v}\mathbf{h}_{\mathrm{u}}^{H}\overline{\mathbf{W}}_{\mathrm{u}}^{H} \mathbf{h}_{\mathrm{u}}}{\mathrm{Tr}\left( \mathbf{U}\overline{\mathbf{W}}_{\mathrm{u}} \right)}
\label{eq55}
\end{aligned}
\end{gather}
\begin{gather}
\begin{aligned}
\mathbf{v}^{H}\frac{\overline{\mathbf{W}}_{\mathrm{t}}\mathbf{F}\overline{\mathbf{W}}_{\mathrm{t}}^{H}}{\mathrm{Tr}\left( \mathbf{F}\overline{\mathbf{W}}_{\mathrm{t}} \right)}\mathbf{v} = \frac{\mathbf{v}^{H}\overline{\mathbf{W}}\mathbf{h}_{\mathrm{f}}\mathbf{h}_\mathrm{f}^{H}\overline{\mathbf{W}}_{\mathrm{t}}^{H}\mathbf{v} }{\mathrm{Tr}\left( \mathbf{F}\overline{\mathbf{W}}_{\mathrm{t}}\right)}
\leqslant \frac{\mathbf{v}^{H}\overline{\mathbf{W}}_{\mathrm{t}}\mathbf{v}\mathbf{h}_{\mathrm{f}}^{H}\overline{\mathbf{W}}_{\mathrm{t}}^{H} \mathbf{h}_{\mathrm{f}}}{\mathrm{Tr}\left( \mathbf{F}\overline{\mathbf{W}}_{\mathrm{t}} \right)}
\label{eq56}
\end{aligned}
\end{gather}
So, $\mathbf{v}^{H}(\overline{\mathbf{W}}_{\mathrm{u}}-\widetilde{\mathbf{W}}_{\mathrm{u}})\mathbf{v}\geqslant 0$ and $\mathbf{v}^{H}(\overline{\mathbf{W}}_{\mathrm{t}}-\widetilde{\mathbf{W}}_{\mathrm{t}})\mathbf{v}\geqslant 0$ hold for any $\mathbf{v} \in \mathbb{C} ^{N_t \times 1}$, which means $\overline{\mathbf{W}}_{\mathrm{u}}-\widetilde{\mathbf{W}}_{\mathrm{u}}$ and $\overline{\mathbf{W}}_{\mathrm{t}}-\widetilde{\mathbf{W}}_{\mathrm{t}}$ are positive semidefinite matrix. It therefore follows that 
\begin{gather}
\begin{aligned}
\widetilde{\mathbf{R}}_{\mathbf{W}}-\widetilde{\mathbf{W}}_{\mathrm{u}}-\widetilde{\mathbf{W}}_{\mathrm{t}}
\succeq 
\overline{\mathbf{R}}_{\mathbf{W}}-\overline{\mathbf{W}}_{\mathrm{u}}-\overline{\mathbf{W}}_{\mathrm{t}}
\succeq 0
\label{eq57}
\end{aligned}
\end{gather}
The constraint (\ref{eq35}e) is satisfied. Above all, the proof is complete.

\balance

\end{document}